\documentclass[twocolumn]{aastex701}
\usepackage{textcomp}
\usepackage{graphicx}
\usepackage{subcaption} 
\graphicspath{{./}}
\usepackage{gensymb}
\usepackage{xcolor}
\usepackage{amsmath}

\begin{document}

\title{Rate of caustic crossing microlensing events in stellar binary lenses with significant orbital motion}

\author[orcid=0009-0004-1245-092X]{Arjun Murlidhar}
\affiliation{Department of Astronomy, The Ohio State University, 140 W 18th Ave, Columbus, OH 43210, USA }
\email{murlidhar.4@osu.edu}  

\author[orcid=0000-0003-0395-9869]{B. Scott Gaudi} 
\affiliation{Department of Astronomy, The Ohio State University, 140 W 18th Ave, Columbus, OH 43210, USA }
\email{gaudi.1@osu.edu}

\author[0000-0003-2377-9574]{Todd. A. Thompson} 
\affiliation{Department of Astronomy, The Ohio State University, 140 W 18th Ave, Columbus, OH 43210, USA }
\email{thompson.1847@osu.edu}

\begin{abstract}

Binary lens microlensing events in which the source crosses a caustic produce sharp, distinctive magnification peaks and can therefore be readily identified. In this paper, we explore the importance of binary orbital motion for the binary lens caustic crossing cross section. If the orbital timescale of the binary system is smaller than the Einstein ring crossing timescale, the caustics sweep out a larger area in the source plane, generally enhancing the cross section. We find that face-on binaries in circular orbits exhibit a substantial increase (up to 4$\times$) in the cross section for caustic crossings. However, highly inclined orbits produce a net decrease relative to a static face-on binary with the same semi-major axis. Using a sample of $\sim 2800$ synthetic binary microlensing events drawn from a realistic Milky Way population-synthesis model, we calculate the average change in the caustic crossing cross section for individual systems with and without orbital motion. Although orbital motion can significantly alter the cross section for specific geometries, we find that, when averaged over the full population, it produces only a small, negligible increase of $\sim 0.1\%$ in the average cross section. We also compute the overall rate of caustic crossing binary events in the simulated sample and find that, with sufficiently dense photometric sampling, $6.3 \pm 0.2\%$ of microlensing events with impact parameter $u_0 < 1$ should exhibit caustic crossings. This rate depends on the distributions of binary separation, mass ratio, and multiplicity of stars and compact objects, and can therefore be used to test our understanding of these underlying properties.
\end{abstract}


\section{Introduction}
\label{intro}
Gravitational microlensing occurs when a star (lens) passes very close to our line of sight to a more distant background star (source), causing a temporary increase in brightness (magnification) of the background source star. When the source and lens can be considered as single point sources, there is a smooth and symmetric increase and decrease in brightness, described by a simple formula \citep{Paczynski1986}. 

But, a significant fraction of stars in the galaxy exist in binary systems. Binary-lens microlensing events can contain additional perturbations and asymmetries on top of the single-lens light curve. The most distinctive of these are caused by \textit{caustic crossing} events, which exhibit sharp, short-timescale spikes in magnification. Caustics are closed curves in the source plane along which the magnification of a point source diverges, and that of a real, finite-sized source is very high \citep{Pejcha2009}. Caustic crossing events are easily identifiable and provide tighter constraints on binary parameters. For surveys with reasonably high cadence and continuous light curve coverage, the sample of caustic crossing binaries will be nearly complete. Thus, the rate of caustic crossing events is a robust, nearly detection efficiency independent metric that can be used to infer distributions of various binary star system parameters.

The Nancy Grace Roman Space Telescope (Roman) will conduct a microlensing survey in the galactic bulge (Galactic Bulge Time Domain Survey or GBTDS) and is expected to detect on the order of 30-50 thousand microlensing events during its 5-year mission and more than a thousand exoplanets across all mass scales \citep{Penny2019}. Roman will also detect thousands of stellar binaries and will be sensitive to isolated and binary compact objects. Since the detectability of a binary system in microlensing depends only on the total system mass (and projected separation) and not on the luminosity of components of the system, Roman will be able to constrain the demographics of non-interacting remnant binaries as well as very low-luminosity stellar and brown-dwarf binaries in the galaxy (e.g., \citet{Choi2013, Malpas2022}).

\citet{mao1991} estimated that $\sim 7 \%$ of all binary star microlensing events will be caustic crossing. However, caustic crossing events dominate the sample of currently known binary lens events because of their enhanced detectability. Binary lens event catalogs published by ground based microlensing surveys MACHO and OGLE between 2000 and 2008 found that $1.6-6\%$ of all their events contain a binary lens, and $\sim 65 - 100 \%$ of these binary lens events are caustic crossing \citep{Alcock2000a, Jaroszynski2002, Jaroszynski2004, Jaroszynski2006, Skowron2007, Jaroszynski2010}. Detailed simulations of microlensing in the galactic bulge show that $23 \%$ of all microlensing events have a binary star lens \citep{Abrams2025}, implying that most of the non-caustic crossing binary lens events were missed by these early microlensing surveys. Roman's superior photometric precision and light curve coverage is expected to increase the fraction of non-caustic crossing events detected; however, caustic crossing events would still represent a significant portion of all events.

Theoretical calculations and simulations of the event rate of binary lens events (caustic crossing or otherwise) to date have assumed the lens to be static throughout the event. However, this is not strictly true, since we routinely find binary microlensing events that cannot be fit by a static binary lens model.  In some cases, static binary lens fits reproduce the overall structure of the event but nevertheless result in subtle yet significant residuals (e.g., \citet{An2002, Han2024}). In other cases, there exist additional large perturbations or caustic crossings that cannot be reproduced by a static lens (e.g., \citet{Albrow2000, Gaudi2008, Bennett2010}). If the orbital period of binary stars is a significant fraction of the time scale of the event, $t_{\rm E}$, then the rate of caustic crossing events could be different for such lenses. 

\begin{figure*}[ht!]
    \centering
    \includegraphics[width=0.95\linewidth]{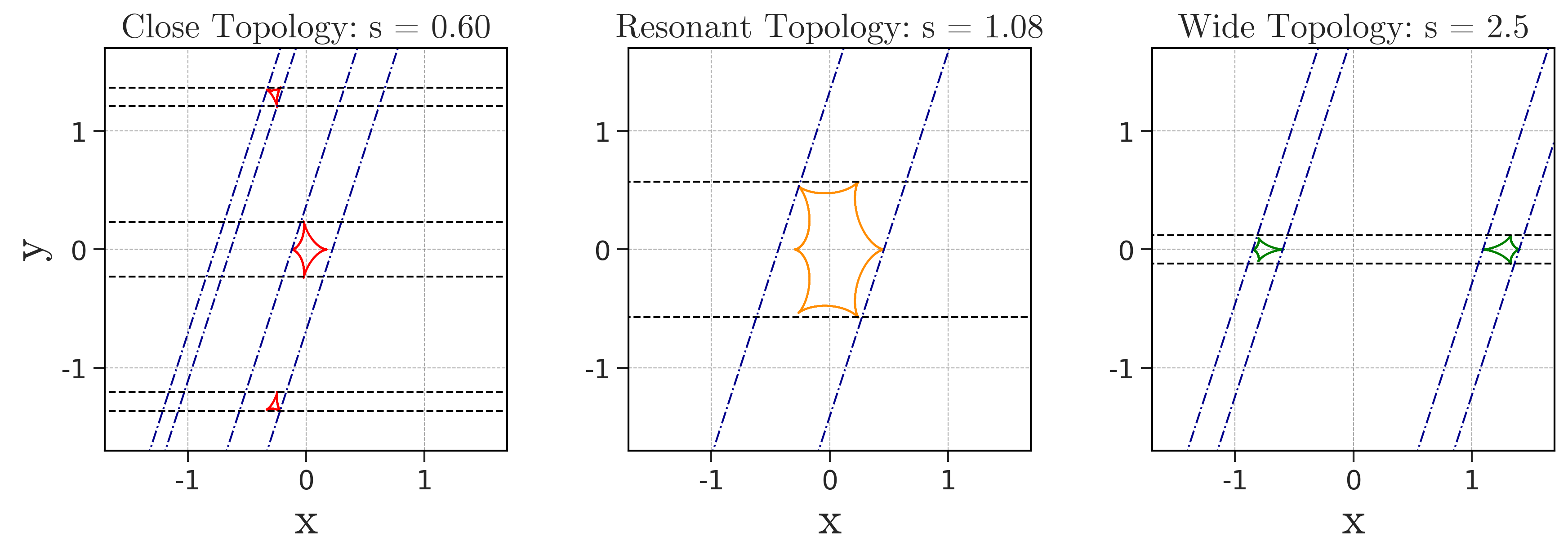}
    \caption{An example of the cross section calculation for close (red curves), resonant (orange curve), and wide (green curves) caustics for $q$ = 0.6. The center of mass of the binary lens is located at the origin, and both lens components lie on the X-axis, with the more massive companion positioned to the left of the origin. The black dashed lines show the width of the caustics for a source trajectory angle $\alpha = 0 \degree$, and blue dots and dashes show the caustic widths for $\alpha = 72 \degree$. The total width for any $\alpha$ is the sum of the individual widths of each disjoint caustic component. For example, in the close topology figure, the total width for $\alpha = 0 \degree$ is the sum of the widths of 3 separate caustic structures and that for $\alpha = 72 \degree$ is the sum of the widths of 2 disjoint components. The caustic widths when $\alpha = 0 \degree$ for the close, resonant, and wide topologies are 0.7767, 1.1443, and 0.2395 respectively, and when $\alpha = 72 \degree$ they are 0.4501, 0.8482, and 0.5321 respectively.}
    \label{fig:caustictopologies}
\end{figure*}

The goal of this paper is to investigate whether orbital motion changes the rate of caustic crossing events significantly, and to quantify this change for different orbital periods of the binary. In Section \ref{sec2}, we discuss caustic structures of static binary lenses and the methodology used in this study to calculate the rate of caustic crossing events for static and orbiting binaries. We show that the rate of caustic crossings is proportional to the average ``cross section'' of caustics (\S \ref{sec2.2}). In Section \ref{sec3}, we calculate the caustic cross section of static binaries in order to compare with the cross sections for binaries with orbital motion. In Section \ref{sec4}, we find how the cross sections change when orbital motion is included and in Section \ref{sec5}, we calculate the total change in cross section for a realistic sample of binary star microlensing events drawn from a galactic population synthesis model. 

\section{Methodology}
\label{sec2}
\subsection{Binary lens equation}
\label{sec2.1}
The lens equation for a binary point-mass lens maps the angular position of the images to the angular position of the source for a given lens configuration. It can be written in complex coordinates as \citep{Witt1990}

\begin{equation}
    \label{lensequation}
    z_{\mathrm{s}}=z-\frac{1}{1+q}\left(\frac{1}{\bar{z}-\bar{z}_1}+\frac{q}{\bar{z}-\bar{z}_2}\right) \, ,
\end{equation}
where $z_{\mathrm{s}}$ is the position of the source in complex coordinates ($z = x + iy$), $z$ is the position of the image, $z_1$ and $z_2$ are the positions of the lens components, and $q = M_2/M_1$ is the mass ratio between the lens components. All angular positions are scaled to the angular Einstein ring radius ($\theta_{\rm E}$), which is the typical angular scale of a microlensing event,

\begin{equation}
    \theta_{\rm E} = \sqrt{\kappa M \pi_{\rm rel}} \, ,
    \label{eqn:thetae}
\end{equation}
where $\kappa = 4G/c^2 {\rm AU} \simeq 8.144\ \  {\rm mas}\ M_\odot^{-1}$, $M_{\rm tot}$ is the total mass of all the lens components, $\pi_{\rm rel} = \pi_{\rm l} - \pi_{\rm s}$ is the relative lens-source parallax, and the lens and source parallaxes have their usual definitions of $\pi_{\rm s} = {\rm au}/D_{\rm s}$ and  $\pi_{\rm l} = {\rm au}/D_{\rm l}$, and $D_{\rm s}$ and $D_{\rm l}$ are the distances to the lens and source, respectively. For a binary lens, the lens equation is equivalent to a fifth order complex polynomial in $z$, which has 5 roots.  Depending on the location of the source with respect to the lenses, either 3 or 5 of these roots actually solve the lens equation and thus are real images.  The magnification of each image $z_j$ is given by the inverse of the determinant of the Jacobian of the mapping given by Eq \ref{lensequation} evaluated at that image position, 
\begin{equation}
    \mu = \left.\frac{1}{\operatorname{det} J}\right|_{z=z_j}, \, {\rm det}J  \equiv
    1-\frac{\partial z_s}{\partial\bar{z}} \overline{\frac{\partial z_s}{\partial\bar{z}}}.
\end{equation}
The magnification of one or more images diverges when $\operatorname{det} J = 0$. The set of all image positions where this happens form closed \textit{critical curves}, and the corresponding set of source positions define closed non-intersecting curves called \textit{caustics}. Although the magnification of a point source diverges along caustic curves, that of a real, finite-sized source is large but finite \citep{Pejcha2009}. Caustics also form boundaries of regions with different numbers of images. For a binary lens, two additional images are formed when a source passes from inside to outside the caustic. The topology of caustics depends on the projected separation of the lens components in units of the Einstein ring radius ($s$) and their mass ratio ($q$), and is classified into three types: \textit{close}, \textit{resonant/intermediate}, and \textit{wide}. For a fixed $q$, the boundaries of these topologies are given by \citep{Schneider:1986,Dominik:1999},

\begin{equation}
    \frac{q}{(1+q)^2}=\frac{\left(1-s_{\rm c}\right)^3}{27 s_{\rm c}^8}, \, s_{\rm w}=\frac{\left(1+q^{1 / 3}\right)^{3 / 2}}{(1+q)^{1 / 2}} \,  .
\end{equation}
If $s \le s_{\rm c}$ (close topology), there is one diamond-shaped caustic near the center of mass of the binary lens and two triangular caustics; if $s_{\rm c} \le s \le s_{\rm w}$ (resonant topology), there is one large caustic near the center of mass; and if $s \ge s_{\rm w}$ (wide topology) there are two diamond-shaped caustics. Fig. \ref{fig:caustictopologies} shows the caustics in these three regimes for a stellar binary lens.

\subsection{Rate of caustic crossing events}
\label{sec2.2}
The rate of microlensing events is defined as the rate at which a background source undergoes microlensing due to a foreground lens. If we assume that the minimum angular separation between the lens and the source for a microlensing event to be detectable is $u_{0,\rm max} \theta_{\rm E}$, then the probability that a given source will exhibit a microlensing event due to a given lens in a time $dt$ is the fraction of the sky covered by the solid angle of width $2u_{0,\rm max} \theta_{\rm E}$ and length $\mu_{\rm rel} dt$. Thus, the contribution of any lens-source pair to the total event rate ($\Gamma_{\rm tot}$) is,

\begin{equation}
    \Gamma_{\rm i}  \propto 2 \mu_{\rm rel, i} u_{0,{\rm max}} \theta_{\rm E, i} 
    \label{eqn:gamma}
\end{equation}
where  $\mu_{\rm rel, i}$ is the relative lens-source proper motion and $\Gamma_{\rm tot} = \sum_{\rm i} \Gamma_{\rm i}$. To find the probability of microlensing of a particular source, this expression is integrated over all lenses along the line of sight to the source (For a more detailed explanation, see \citet{gaudi2012, Griest91, Penny2013}). One can similarly define the contribution of an event to the rate of caustic crossing events as 

\begin{equation}
    \Gamma_{ \rm cc, i}  \propto  \mu_{\rm rel, i} w_{\rm i}  \theta_{\rm E, i} 
    \label{eqn:gammacc}
\end{equation}
where $ w_{\rm i}$ is the average width of the caustic in units of $\theta_{\rm E, i}$ (See \S 2.3 and Fig. \ref{fig:caustictopologies}). We also refer to this average width as the (linear) cross section to caustic crossings. Therefore, for a given event $i$, the probability of it being a caustic crossing event is $w_{\rm i}/2u_{\rm 0, max}$. The fraction of all events that have caustic crossings is $\langle w \rangle/2u_{0,{\rm max}}$, where $\langle w \rangle$ is the cross section averaged over all systems, including single lenses (which do not produce caustic crossings), and binary lenses (which do). Thus, $\langle w \rangle$ is proportional to the rate of caustic crossing events. If $f_{\rm bin}(M)$ is the binary fraction as a function of mass, then $\langle w \rangle$ is the weighted average cross section over all binary lens systems using $f_{\rm bin}(M)$ as the weight for each system. 

\subsection{Calculating caustic cross sections}
\label{sec2.3}


We define the cross section of a caustic structure as the angle-averaged ``angular width'' of the caustic in units of the Einstein radius. The angular width of a caustic for any particular source trajectory angle, $\alpha$, is the range of impact parameters for which the source trajectory crosses the caustic. If there are multiple disjoint caustic components, then the total width for any $\alpha$ is the sum of the widths of each caustic component. Fig. \ref{fig:caustictopologies} shows how the caustic width is calculated for different source trajectory angles. 

We calculated caustic cross sections for different values of the binary lens projected separation $s$, and mass ratios $q$. For each $(s, q)$ configuration, and for a given source-trajectory angle $\alpha$ (measured with respect to the binary axis), we simulated between 2000-10000 parallel source trajectories depending on the value of $s$ and measured the transverse separation between the first trajectory that enters a region bounded by a caustic and the last trajectory that is inside this region. This separation is the width of the caustic for the given angle $\alpha$. Source trajectories were limited to a maximum distance of $\beta\theta_{\rm E}$ from the origin (which we define as the center-of-mass of the lens), where $\beta$ depends on the value of $s$. Points were uniformly sampled along the trajectory between $t_{\rm start} = t_0 - \beta t_{\rm E}$ and $t_{\rm end} = t_0 + \beta t_{\rm E}$ at intervals of $10^{-3}t_{\rm E}$, where $t_{\rm E}$ is the timescale of the microlensing event and is defined as the time taken for a source to cross the Einstein ring,

\begin{equation}
    t_{\rm E} = \frac{\theta_{\rm E}}{\mu_{\rm rel}}.
    \label{eqn:te}
\end{equation}

Thus, for a typical timescale of $t_{\rm E} \sim 15~{\rm days}$, the sampling interval corresponds to $\sim 22$ minutes. Here $t_0$ is the time of the closest lens-source separation and the projected lens-source distance at this time in units of $\theta_{\rm E}$ is called $u_0$. Whether or not any part of the trajectory was inside the caustic was determined by counting the number of images at every point along the trajectory using \textit{MulensModel} \citep{mulensmodel2, mulensmodel}, which employs \textit{VBBLensing} \citep{bozza2010} and the root-finding algorithm by \citet{skowron&gould2012} to solve the lens equation. The spacing between trajectories was also set at $10^{-3} \theta_{\rm E}$. The source plane is scanned sequentially with these trajectories for any given $\alpha$. Every time a trajectory enters or exits a caustic, we perform a binary search to a specified tolerance limit (set at $\sim 10^{-4} \theta_{\rm E}$) to find the caustic boundary with better precision. When we encounter a caustic, we also increase the sampling cadence along the trajectory to match the tolerance limit set for the binary search ($\sim 10^{-4} t_{\rm E}$). We used $\beta = 5$ for $s \leq 0.5$, $\beta = 2$ for $0.5 < s \leq 2$ and $\beta = 3$ for $2 < s \leq 6$. For $s > 6$, we only solve the lens equation for source trajectory points within a torus centered on the origin with $r_{in} = 3\theta_{\rm E}$ and $r_{out}=6\theta_{\rm E}$ to reduce the computation time (this is discussed further in Appendix \ref{A3}). For inclined orbiting binaries, since $s$ changes with time (see Sec. \ref{sec4.2}), we use $\beta = 5$ for all $s$. The values of $\beta$ were chosen to optimize the computation time while ensuring that the numerical errors are not too large. 

If we were interested only in static cross sections, we could calculate them essentially exactly (up to numerical precision). The mean width of a caustic (a closed, concave curve) is equal to the perimeter of the smallest convex curve that encloses it divided by $\pi$ (see Appendix \ref{secA1}). This enclosing curve is known as the {\it convex hull}, and for binary lens caustics, it can be constructed by connecting adjacent cusps (vertices) with straight-line segments. For any binary lens configuration, the critical curves (and hence the caustics) can be determined by solving a quartic complex polynomial \citep{Witt1990}. The cusp positions can then be found by locating the points at which the tangent vector to the caustic vanishes. Thus, the locations of all cusps can be determined to numerical precision.\footnote{In some limiting cases, such as planetary lenses with $q \ll 1$, the cusp positions can be obtained analytically \citep{Bozza1999}.} Once the cusp locations are known, the convex hull and therefore the mean width can be calculated straightforwardly. However, this approach is not practical for orbiting binaries, for which the caustic changes continuously with time. For this reason, we adopted the numerical method described above to compute the cross sections of both static and orbiting binaries.

We calculated the cross section of caustics for any $(s,q)$ by averaging the caustic width over 20 $\alpha$ values uniformly sampled between 0 and $2\pi$. For each value of $s$, the cross section was calculated for 20 uniformly spaced $q$ between 0.05 and 1, and the average cross section over these $q$ values was found. Our numerical calculation can find caustic widths with a precision of $\sim 10^{-3} \theta_{\rm E}$. This precision is sufficient to resolve all caustics in the range of binary lens parameters used in this paper. For $s < 0.2$, the distance from the center of mass to the two triangular caustics is greater than $5\theta_{\rm E}$. Therefore, these caustics are missed by our computation, resulting in an error of $\sim O(10^{-3} \theta_{\rm E}$) (roughly the size of a triangular caustic when $s=0.2$ and $q = 1$) which corresponds to a relative error of $\sim 10 \%$ when $s = 0.2$. For certain combinations of large $s$ ($> 4$) and small $q$ values, one or sometimes both diamond caustics are missed since they fall outside the bounds of our trajectories (see Appendix \ref{A3}), resulting in an average error of $\sim O(10^{-2} \theta_{\rm E})$ (i.e., the order of the size of one of the diamond caustics) in the cross section in this regime. To estimate the error in the cross section due to the finite sampling of $\alpha$, we calculated the average static cross section for four $s$ values averaged over 20 $q$ values for each $s$ and 100 $\alpha$ values instead of 20. We find that the average difference between the two calculations is $\langle w(20 \alpha) - w(100\alpha) \rangle = -0.0048 \theta_{\rm E}$, and the relative error ranges from -0.002 to -0.04 as we move from larger to smaller cross sections. These errors are sufficiently small that they do not impact our conclusions.

This numerical method for calculating cross sections can be easily extended to orbiting lenses. For a general orbiting lens, the angle between the binary axis and the trajectory $\alpha$, and the projected separation $s$ between the lens components will change with time. At each time step in the trajectory, we update $s$ and $\alpha$ to account for orbital motion, and follow the same procedure as the static lens scenario to calculate cross sections. We also need to account for the different orbital phases (evaluated at $t_0$) that the lens can have. For a face-on lens in a circular orbit, sampling different phase values is degenerate with using different $\alpha$ values to compute the cross section. When the lens is inclined, we calculate cross sections for 10 phase ($\phi$) values and average them. 

\section{Caustic cross sections for static lenses}
\label{sec3}
\begin{figure}[ht!]
    \centering
    \includegraphics[width=0.95\linewidth]{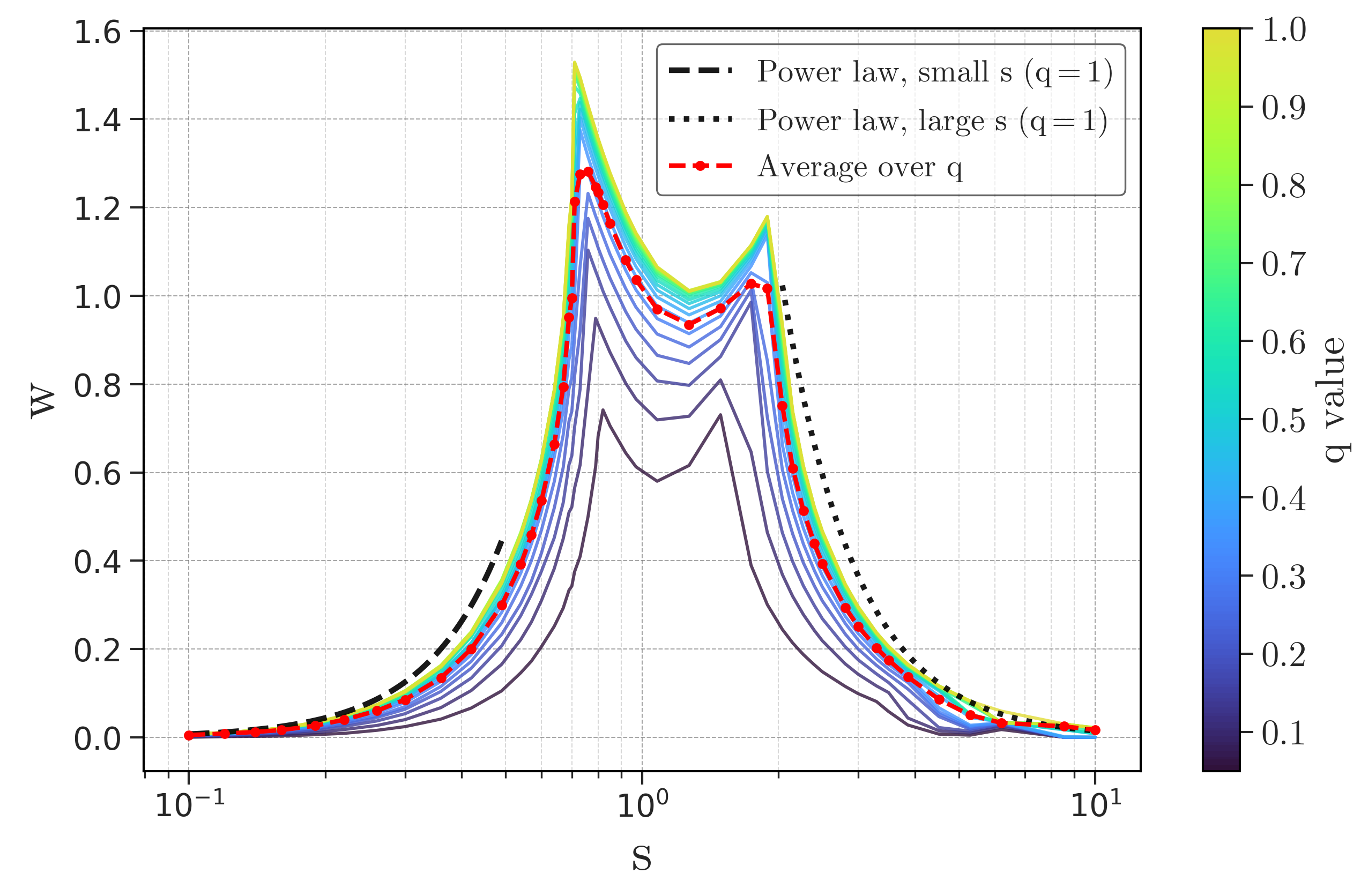}
    \caption{Cross section of caustics for a static binary lens as a function of the separation of lens components (in units of $\theta_{\rm E}$) for 20 $q$ values between 0.05 and 1 plotted using colored lines. The average cross section over these 20 $q$ values is plotted in red. }
    \label{fig:static_crs}
\end{figure}

In Fig. \ref{fig:static_crs}, we have plotted the static cross section as a function of $s$ for 20 uniformly spaced $q$ values between 0.05 and 1.00 with colored lines and the average cross section over all $q$ values in red. The peaks of this distribution occur roughly at the boundaries between close-resonant ($s_{\rm c}$) and resonant-wide ($s_{\rm w}$) topologies. For $q = 1$, this occurs at $s_{\rm c} = \sqrt{2}$ and $s_{\rm w} = 2$ \citep{Schneider:1986}. For small $q$ values, there is a bump near $s \sim 6$ which is an artifact of the numerical errors in our calculation. At $s = 6$, we switch from sampling points along all trajectories with $u_0 < 3$ to only points in a torus as described above. For small $q$ values around $s = 6$, both these sampling strategies miss one of the two diamond caustics. 

\citet{B&G2001} have also calculated the ``mean width'' of caustics as a function of $s$ using an analytic method. While our results are similar to theirs qualitatively, they differ in the actual quantity being computed. The difference between the two calculations is explained in Appendix \ref{secA1}.

The tails of this distribution are fitted well by power laws. For small $s \ (s < 0.5)$, $w \simeq 2.81 \ s^{2.58}  q^{0.50} $ and in the limit of large $s \ (s > 2)$, $w \simeq 6.87 \ s^{-2.67}  q^{0.49} $. To reproduce the \citet{mao1991} calculation of the fraction of caustic crossing events, we integrated the cross section functions in Fig. \ref{fig:static_crs} over $s$ and $q$. We used a log-uniform distribution in $s$ with $\log(s) \in [-2, 4]$ and a uniform distribution in $q \in [0.1, 1.0]$, and extrapolated the cross section function to small and large s using the fitted power laws. We obtained an average cross section of 0.126. The cross section for a microlensing event with $u_0 < 1$ is 2. Therefore, the ratio of the number of caustic crossing events to the number of binary microlensing events with $u_0 < 1$ is  $0.126/2 = 0.063$ (See Eqs. \ref{eqn:gamma} and \ref{eqn:gammacc}). This is consistent with \citet{mao1991}'s value of $0.065$. 

We also calculated the relative rate of caustic crossing events for other $q$ and $s$ distributions, which can be found in Table \ref{tab:caustic-cross-frac}. To find the rate of caustic crossing events relative to all microlensing events, we need to multiply these numbers with the binary fraction (which is calculated in Sec. \ref{sec5.4}).

\begin{table}
    \centering
    \begin{tabular}{|c|c|c|}
        \hline
        & log uniform s & log normal s \\
        &                &             $\mu = -0.57, \ \sigma = 0.88$ \\ \hline
        dN/d$q$ $\propto \text{const} $& $6.30\%$ & $12.9\%$ \\ \hline
        dN/d$q$ $\propto$ $q$  & $6.63\%$ & $13.6\%$ \\ \hline
        dN/d$q$ $\propto$ 1/$q$ & $5.71\%$ & $11.7\%$ \\ \hline
    \end{tabular}
    \caption{Rate of caustic crossing binary events relative to binary microlensing events with $u_0 < 1$ for different s and q distributions. $\log(s)$ ranges from -2 to 4 and $q$ ranges from 0.1 to 1 for all the calculations.}
    \label{tab:caustic-cross-frac}
\end{table}

\begin{figure*}[ht!]
    \centering
    \begin{subfigure}[b]{0.45\linewidth}
        \centering
        \includegraphics[width=\linewidth]{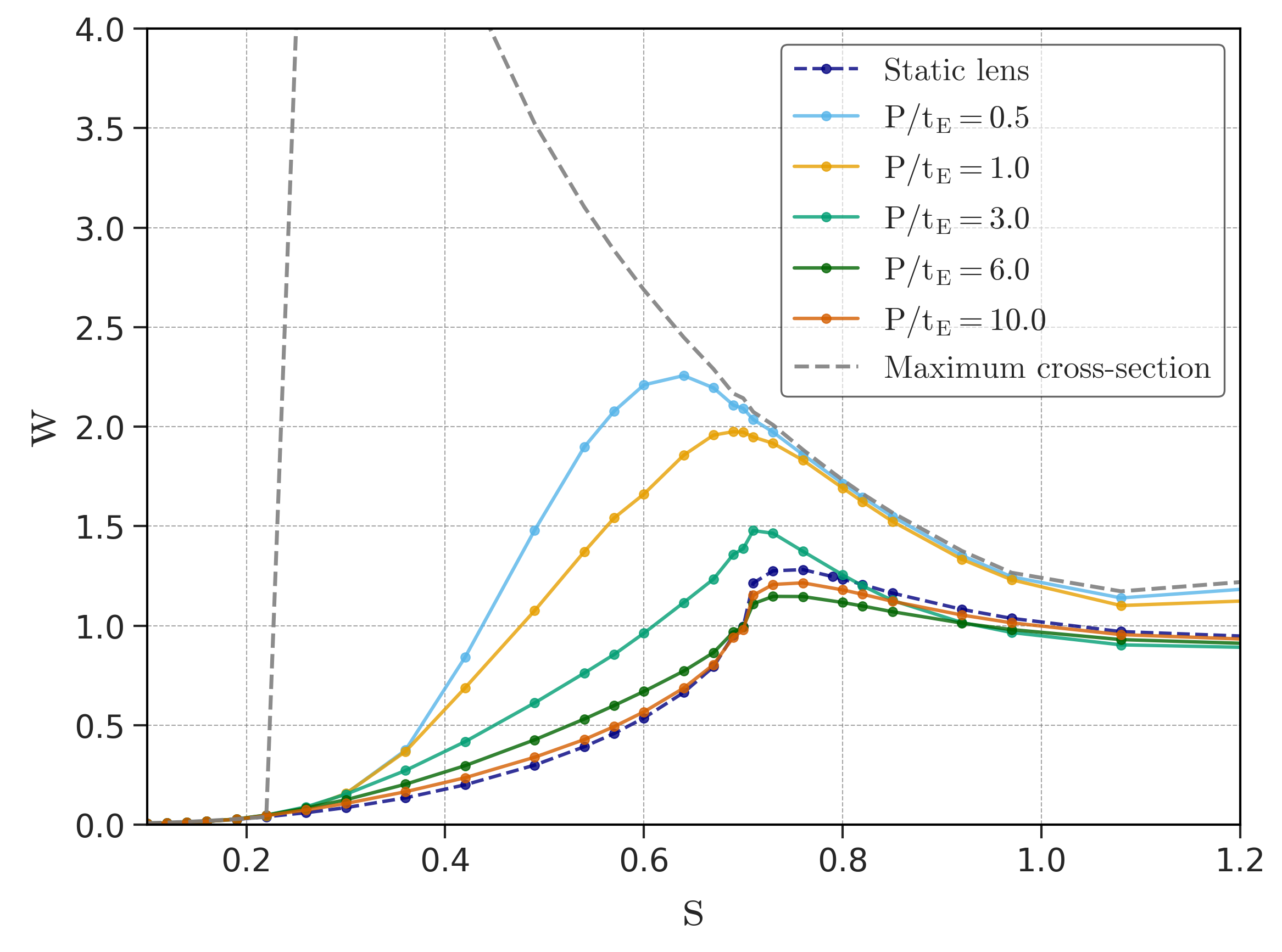}
    \end{subfigure}
    \hfill
    \begin{subfigure}[b]{0.45\linewidth}
        \centering
        \includegraphics[width=\linewidth]{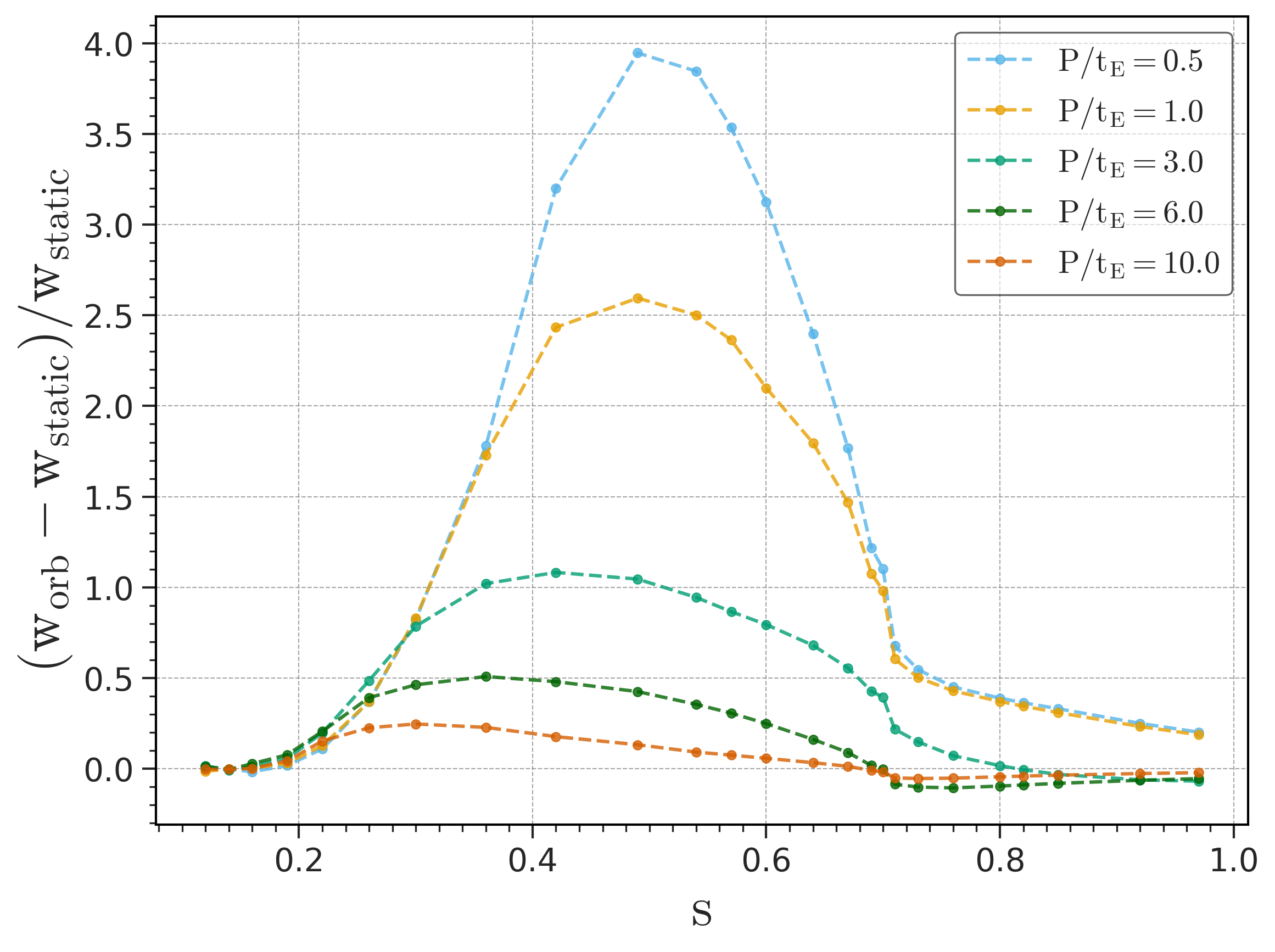}
    \end{subfigure}
    \caption{(Left) Caustic cross section for face-on binary stars in a circular orbit as a function of $s$ for different values of the ratio of orbital period of the binary to the Einstein ring crossing time. Each curve is an average over 20 $q$ values. The gray dashed curve shows the cross section for a face-on binary system with an infinitely small orbital period and is equal to the maximum distance between two points on a caustic curve. (Right) Fractional increase in caustic cross section for different $P/t_{\rm E}$ values with respect to a static lens.}
    \label{Fig:rotation_faceon_cross_sect}    
\end{figure*}

The parameters $\mu$ and $\sigma$ for the log normal distribution of $s$ were derived by fitting a gaussian to the $s$ distribution of a realistic population of binary lens microlensing events produced using the microlensing survey simulation tool \textit{PopSyCLE} \citep{Lam2020}. \textit{PopSyCLE} simulations are discussed further in section \ref{sec5.1}. We can see from Table \ref{tab:caustic-cross-frac} that the average caustic cross section is more sensitive to changes in the projected separation distribution (and hence the semi-major axis distribution) of binaries than to those in the mass ratio distribution.

\section{Caustic cross sections with orbital motion}
\label{sec4}
\subsection{Face-on binary lenses}
\label{sec4.1}
We now examine the effects of binary lens orbital motion by first considering face-on binary star systems in circular orbits. The projected separation of the binary lens system remains constant throughout the orbit, but the inclination of the binary axis with respect to any fixed celestial coordinate system changes constantly. As a result, the caustics also rotate around the center of mass in the source plane. If the orbital period of the binary is on the order of the timescale of the microlensing event, then (in most cases) the rotating caustics sweep out a larger area in the source plane, increasing the probability of a source trajectory encountering one of the caustics. 

\begin{figure*}[ht]
    \centering
    \begin{subfigure}[b]{0.45\linewidth}
        \includegraphics[width=\linewidth]{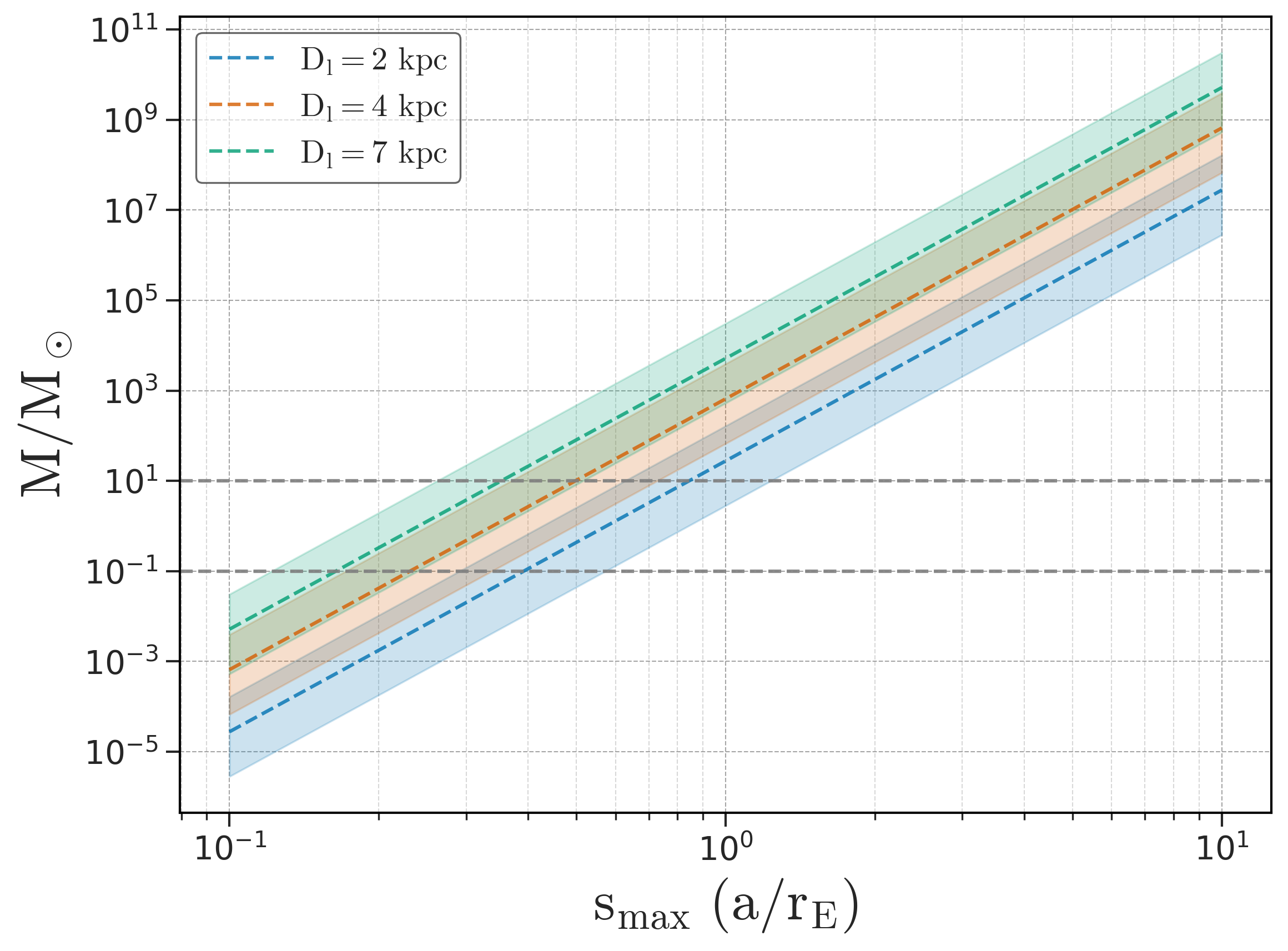}
        \captionsetup{margin={1cm, 0cm}}
        \caption{Primary mass as a function of $s_{\rm max}$ for fixed $P/t_{\rm E} = 10$}
        \label{fig:M_vs_s_Dl}
    \end{subfigure}
    \hfill
    \begin{subfigure}[b]{0.45\linewidth}
        \includegraphics[width=\linewidth]{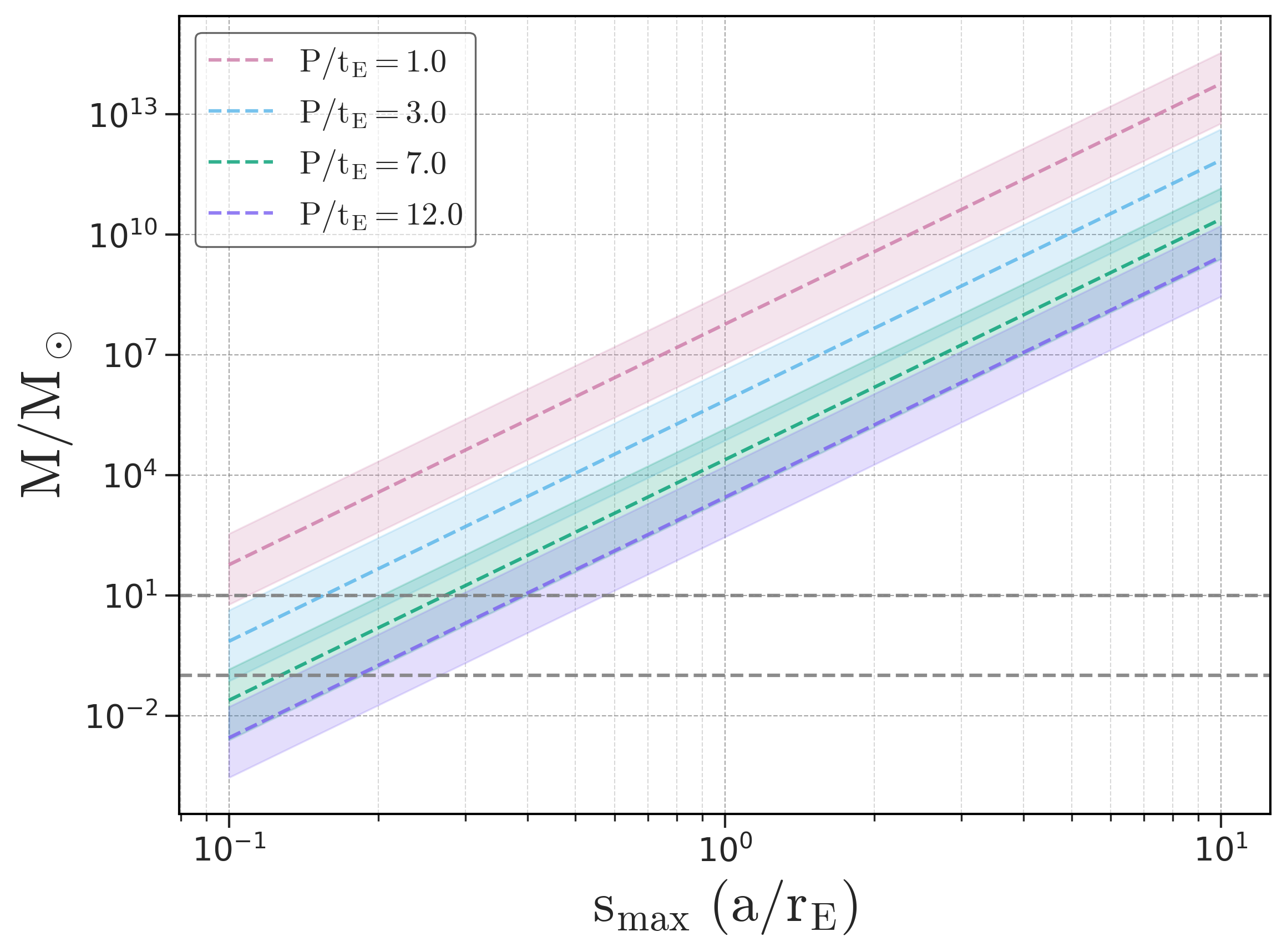}
        \captionsetup{margin={1cm, 0cm}}
        \caption{Primary mass as a function of $s_{\rm max}$ for fixed $D_{\rm l}=7.36 \ Kpc$}
        \label{fig:M_vs_s_Tote}
    \end{subfigure}
    \caption{Plots of primary mass of the binary vs maximum projected separation for different lens distances ($D_{\rm l}$) and $P/t_{\rm E}$ values. The source distance $D_{\rm s}$ is fixed at $9.76 \rm \, kpc$. These relations are obtained by averaging Eq. \ref{eqn:mvss} over a uniform mass ratio ($q$) distribution. The colored dashed lines correspond to the median lens-source relative proper motion value derived from \textit{PopSyCLE} ($\mu_{\rm rel, med} = 6.65 \rm \, mas \, yr^{-1}$), and the shaded regions are the $68\%$ confidence intervals. The two grey dashed lines mark the range of mass values that are most common for stellar or compact object lenses. We expect most binary lenses with $P/t_{\rm E} \leq 10$ to have $s_{\rm max} \lesssim 1$.}
    \label{fig:M_vs_s_plots}
\end{figure*}

Fig. \ref{Fig:rotation_faceon_cross_sect} shows the cross section as a function of $s$ for different values of the ratio between orbital period ($P$) and Einstein ring crossing time $t_{\rm E}$. We restricted the range of $s$ to $\lesssim 1$, since larger $s$ values and small orbital periods would imply very large primary masses which are improbable (see Fig. \ref{fig:M_vs_s_plots}). For $P \lesssim t_{\rm E}$, there is a significant increase in the caustic cross section. For face-on lenses, we can also define a maximum cross section, which is equal to the diameter of the circle centered on the origin that circumscribes the caustics. When the cross section is equal to the maximum cross section for any ($s$,$q$), any trajectory that has an impact parameter smaller than the radius of such a circle will always cross a caustic. The $P/t_{\rm E} = 0.5$ and $ P/t_{\rm E} = 1$ curves approach the maximum cross section for $s$ in the resonant regime. When $P$ is several times the Einstein ring crossing timescale, there is a minimal increase in cross section compared to the static case. When $P/t_{\rm E} = 10$, the largest increase in the cross section is by $\sim 20 \%$ for caustics in the close topology. Another interesting effect we observed is the dip of the orbiting cross section curve below the static curve for $s \gtrsim 0.7$ and $P/t_{\rm E} = 3$ and $P/t_{\rm E} = 6$. This is discussed further in Appendix \ref{A2}.

In order to get a sense of how probable different $P/t_{\rm E}$ values are, we calculated the mean lens mass as a function of $s_{\rm max}$ for different values of $P/t_{\rm E}$ and lens distance $D_{\rm l}$ (Fig. \ref{fig:M_vs_s_plots}). Here, we define $s_{\rm max}$ as the semi-major axis of the binary scaled to the Einstein ring radius $(a/r_{\rm E})$, instead of the instantaneous projected separation. This is a better metric to characterize rapidly orbiting lenses since, in general, the projected separation will change considerably during the microlensing event when $P \sim t_{\rm E}$. For a face-on binary in a circular orbit, $s_{\rm max}$ is the same as the projected separation, $s$, at any point in time. For an inclined circular orbit, it is the maximum projected separation during an orbit.  Using the definitions of $t_{\rm E}$, $\theta_{\rm E}$ (Eq. \ref{eqn:thetae}), and Kepler's third law, and defining,
\begin{equation}
    s_{\rm max} = \frac{a}{D_{\rm l}\theta_{\rm E}} ,
\end{equation}
we get, 
\begin{equation}
\begin{split}
M_{\rm p} &= \left(\frac{16 \kappa \pi^4}{G^2}\right)
       \frac{\pi_{\rm rel}\,\mu_{\rm rel}^4\,(s_{\rm max} D_{\rm l})^6}
            {(P/t_{\rm E})^4\,(1+q)} \\
&\simeq 0.43\,M_\odot
   \left(\frac{\pi_{\rm rel}}{0.145~{\rm mas}}\right)
   \left(\frac{\mu_{\rm rel}}{7~{\rm mas~yr^{-1}}}\right)^{4}
   \left(\frac{s_{\rm max}}{0.3}\right)^{6} \\
&\quad\times
   \left(\frac{D_{\rm l}}{4~{\rm kpc}}\right)^{6}
   \left(\frac{P/t_{\rm E}}{10}\right)^{-4}
   \left(\frac{1+q}{2}\right)^{-1}
\label{eqn:mvss}
\end{split}
\end{equation}
Fig. \ref{fig:M_vs_s_plots} plots $M_{\rm p}$ as a function of $s_{\rm max}$ for different values of $D_{\rm l}$ and $P/t_{\rm E}$. For $D_{\rm s}$ and $\mu_{\rm rel}$, we use median values (and the $68\%$ confidence interval for $\mu_{\rm rel}$) of their distributions derived from a realistic population of binary star microlensing events produced using \textit{PopSyCLE} (See Sec \ref{sec5.1}).
In the mass range of $0.1 - 10 \ M_\odot$, which contains most stars and compact objects in our galaxy, systems with $P/t_{\rm E}$ values where we expect to see a significant effect due to orbital motion are restricted to small $s_{\rm max}$ values. It is highly unlikely that a binary system with $s_{\rm max} > 1$ shows a large change in its caustic cross section due to orbital motion.

From Fig. \ref{fig:M_vs_s_Dl}, we can see that at $P/t_{\rm E} = 10$, an $s_{\rm max}$ greater than 1 can only be achieved with a very small lens distance ($D_{\rm l} \lesssim 2$ kpc), large primary mass ($M_{\rm p} \gtrsim 10 \ M_\odot$) and a small lens-source proper motion. Therefore, we restrict our cross section calculations in this section to $s \lesssim 1$. 

\subsection{Effect of inclination}
\label{sec4.2}
Most stellar binary microlenses we observe will not be face-on, and will instead have an orbit that is inclined with respect to the plane perpendicular to our line of sight. In this case, not only will the binary axis rotate in the plane, but the projected separation between the two stars will also change during the orbit. 


\begin{figure*}[ht!]
    \centering
    \begin{subfigure}[b]{0.45\linewidth}
         \centering
         \includegraphics[width=0.98\linewidth]{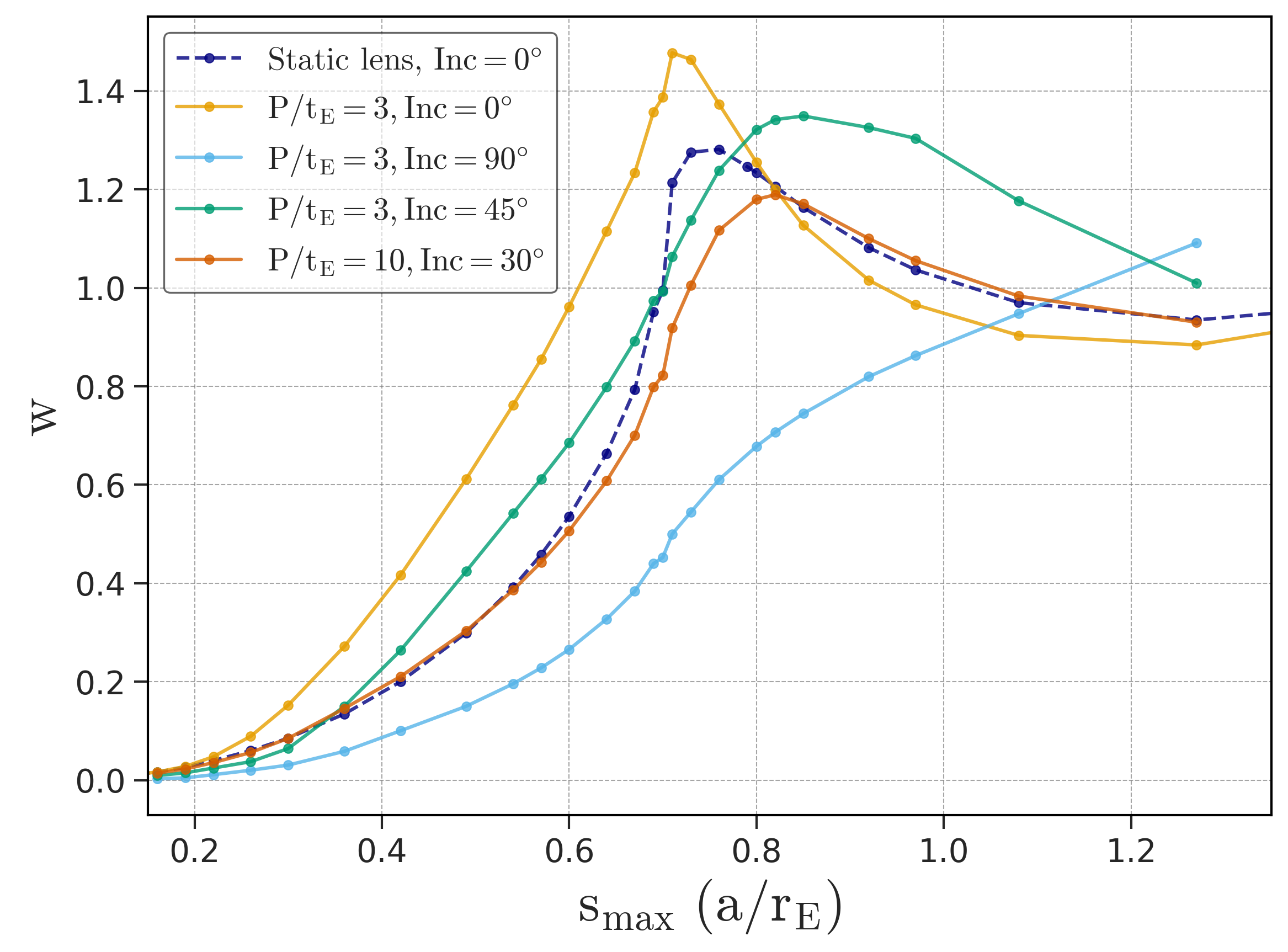}
    \end{subfigure}
    \hfill
    \begin{subfigure}[b]{0.45\linewidth}
         \centering
         \includegraphics[width=0.98\linewidth]{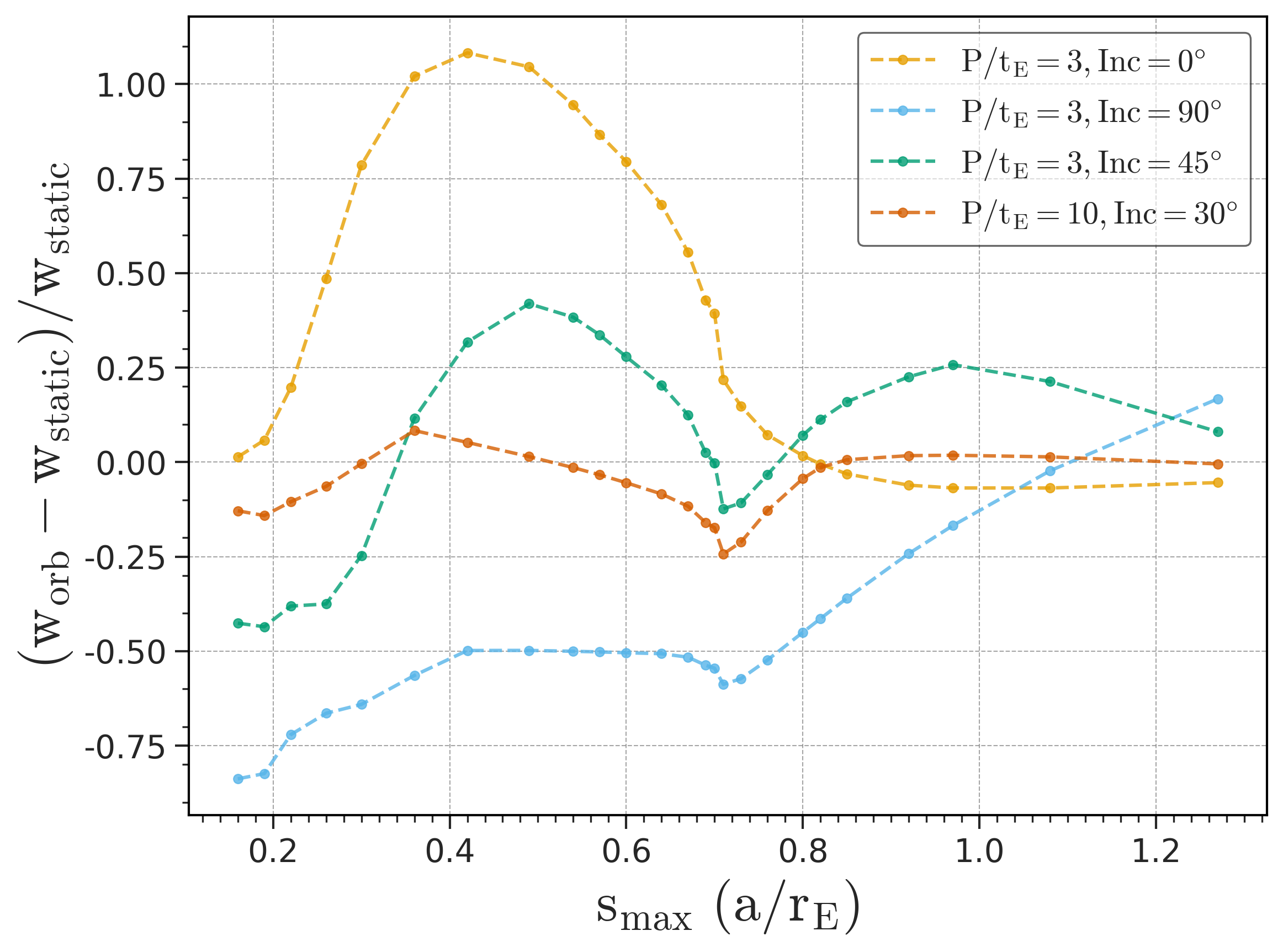}
    \end{subfigure}
    \caption{(Left) Caustic cross section as a function of $s_{\rm max} \ (a/r_{\rm E})$ for stellar binary lenses in circular orbits inclined to our line of sight at different angles. Each point is the average cross section over 20 q values and 10 values of orbital phase $\phi$. This plot compares the cross sections for 3 different inclinations when $P/t_{\rm E} = 3$, and shows the cross section for one value of the inclination when $P/t_{\rm E} = 10$. Binaries on inclined orbits have smaller cross sections than their face-on counterparts. Highly inclined orbiting binaries can also have cross sections smaller than a static face-on binary. (Right) Fractional change in cross section relative to a static face-on lens for the $P/t_{\rm E}$ and inclination values considered in the figure on the left.}
    \label{fig:cross_sect_inc}
\end{figure*}

Fig. \ref{fig:cross_sect_inc} shows how the cross section changes with inclination. Here, an inclination of $0 \degree$ corresponds to a face-on orbit and $90 \degree$ corresponds to an edge-on orbit. Along with averaging over $q$ and $\alpha$, each point in this plot (for the curves with orbital motion) is also the average of the cross section over 10 uniformly sampled orbital phase values ($\phi$). The static lens curve (dark blue) is the cross section for a face-on static lens, i.e., $s_{\rm max} = s$ (projected separation during the event). 

We can see that for $P/t_{\rm E} = 3$, higher inclinations result in smaller cross sections, and an edge-on orbit leads to cross sections that are smaller than the static lens scenario for all $s_{\rm max}$ in the close topology. In an inclined circular orbit, the projected separation between the lenses is equal to the semi-major axis at only two points in the orbit, and is smaller for all other points in the orbit. If $s_{\rm max}$ is in the close topology regime, then for all $s < s_{\rm max}$, the caustics are smaller in size, leading to smaller cross sections. As the inclination increases, the average projected separation during an orbit decreases, and hence the cross section decreases. For larger $P/t_{\rm E}$ values (e.g. see $P/t_{\rm E} = 10$ in Fig. \ref{fig:cross_sect_inc}), even moderate inclinations can negate any increase in cross section due to the rotation of caustics and result in cross sections similar to or smaller than the static case. This trend is reversed in the resonant topology regime because smaller projected separations lead to slightly smaller caustics but more elongated caustic structures and therefore an effective increase in cross section. 

\section{Caustic cross sections for a realistic population of lenses}
\label{sec5}
\subsection{PopSyCLE simulations} 
\label{sec5.1}
To determine whether orbital motion will significantly alter caustic crossing event rates in a real microlensing survey, we used \textit{PopSyCLE} (\citet{Abrams2025, Lam2020}) to run Milky Way microlensing survey simulations and calculated cross sections for the population of binary lens microlensing events produced. \textit{PopSyCLE} performs a population synthesis of the Milky Way with a combination of single stars, compact objects, and multiple systems. We can then carry out a mock microlensing survey over any number of galactic fields for a given duration of time using \textit{PopSyCLE}. 

The sample of binary lens events used for our calculation was constructed in two steps. First, we used simulations produced by \citet{Abrams2025} and selected all binary lens microlensing events. Abrams et al ran \textit{PopSyCLE} simulations for 18 galactic bulge fields, each with an area of 0.34 deg$^2$, from the OGLE IV survey. This survey was run for 1000 days and candidate microlensing events were identified using observability criteria to match the limits of OGLE observations. The cuts used were $I_{\rm base} \leq 21$, I-band bump magnitude $\geq 0.1$, and $|u_0| \leq 2$. $I_{\rm base}$ is the baseline magnitude in the I-band, and the bump magnitude ($\Delta m$) is the difference between the baseline and peak magnitudes of the lightcurve. This yielded $\sim 1100$ binary lens microlensing events.

\begin{figure*}[ht!]
     \centering
     \begin{subfigure}[b]{0.48\textwidth}
         \includegraphics[width=\textwidth]{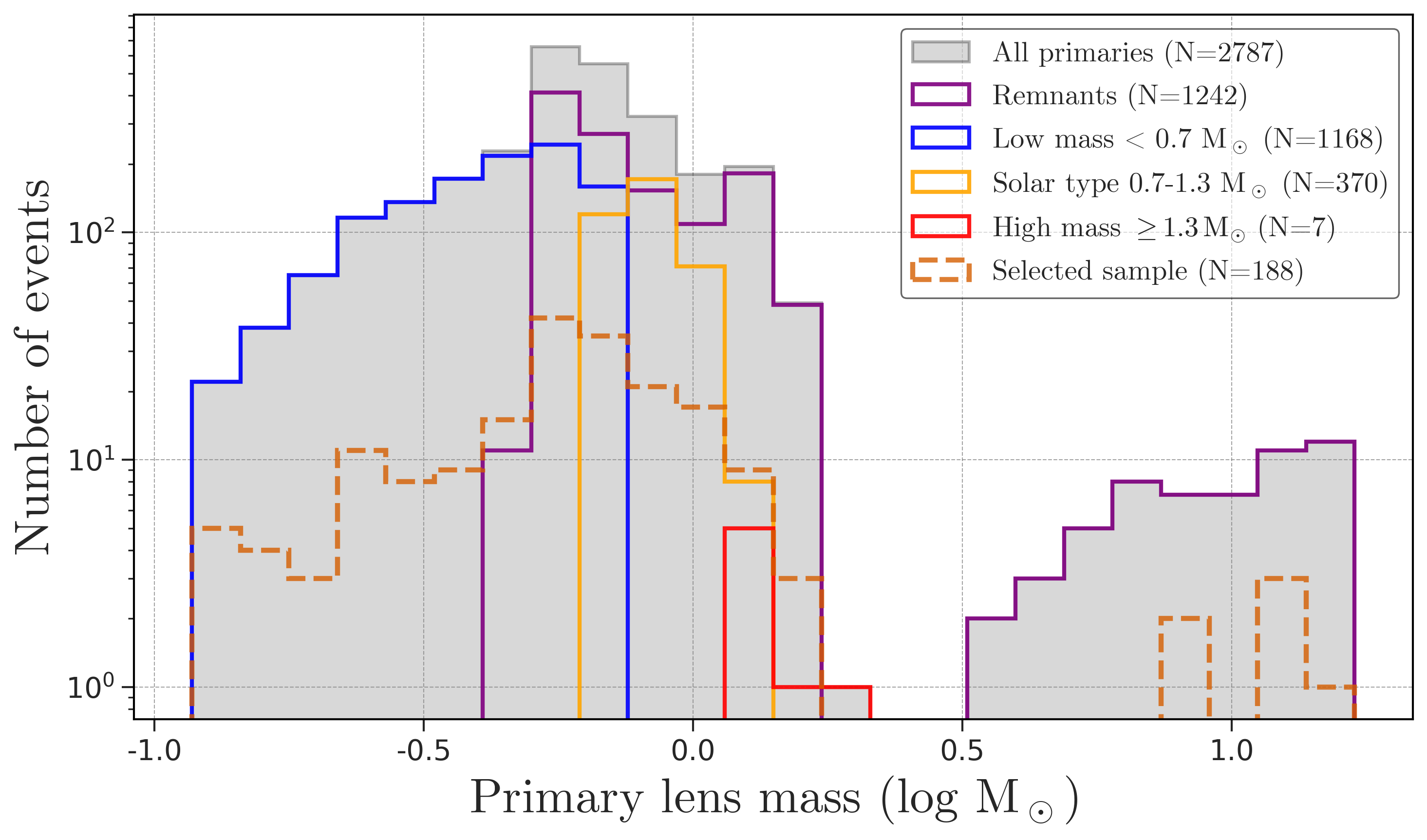}
         \caption{Primary lens mass distribution of all binary lens events (gray shaded) with contributions from different types of primary lenses shown with colored histograms. The dashed line histogram shows the sample of events we selected for the cross section calculation.}
         \label{fig:mass_dist_full}
     \end{subfigure}
     \hfill
     \begin{subfigure}[b]{0.44\textwidth}
         \includegraphics[width=\textwidth]{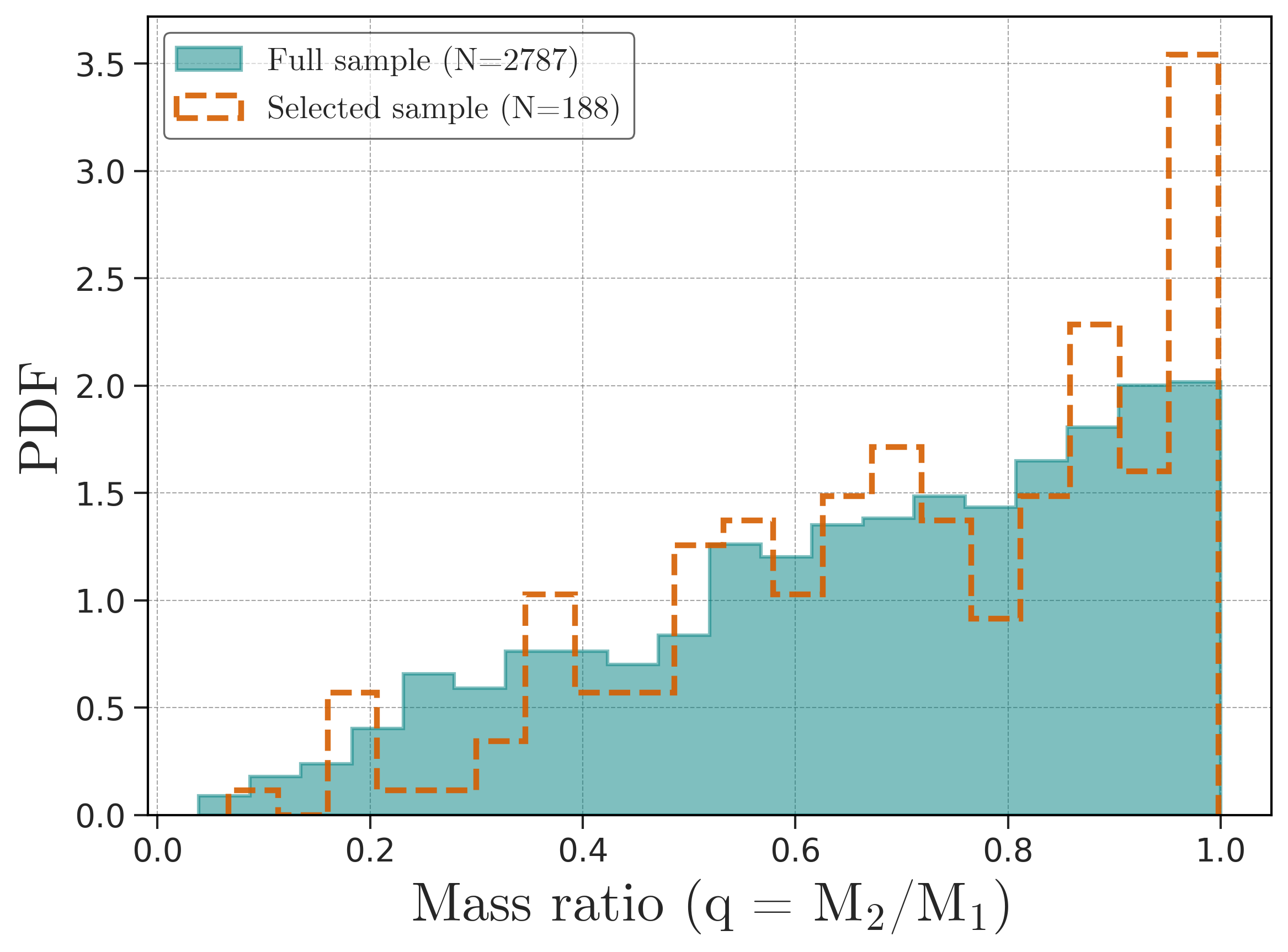}
         \caption{Probability density function of the mass ratio between the lens components for all the binary lens events (cyan) and the selected subset (dashed line).}
         \label{fig:mass_ratio_dist}
     \end{subfigure}

     \vspace{10pt} 

     \begin{subfigure}[b]{0.55\textwidth}
         \includegraphics[width=\textwidth]{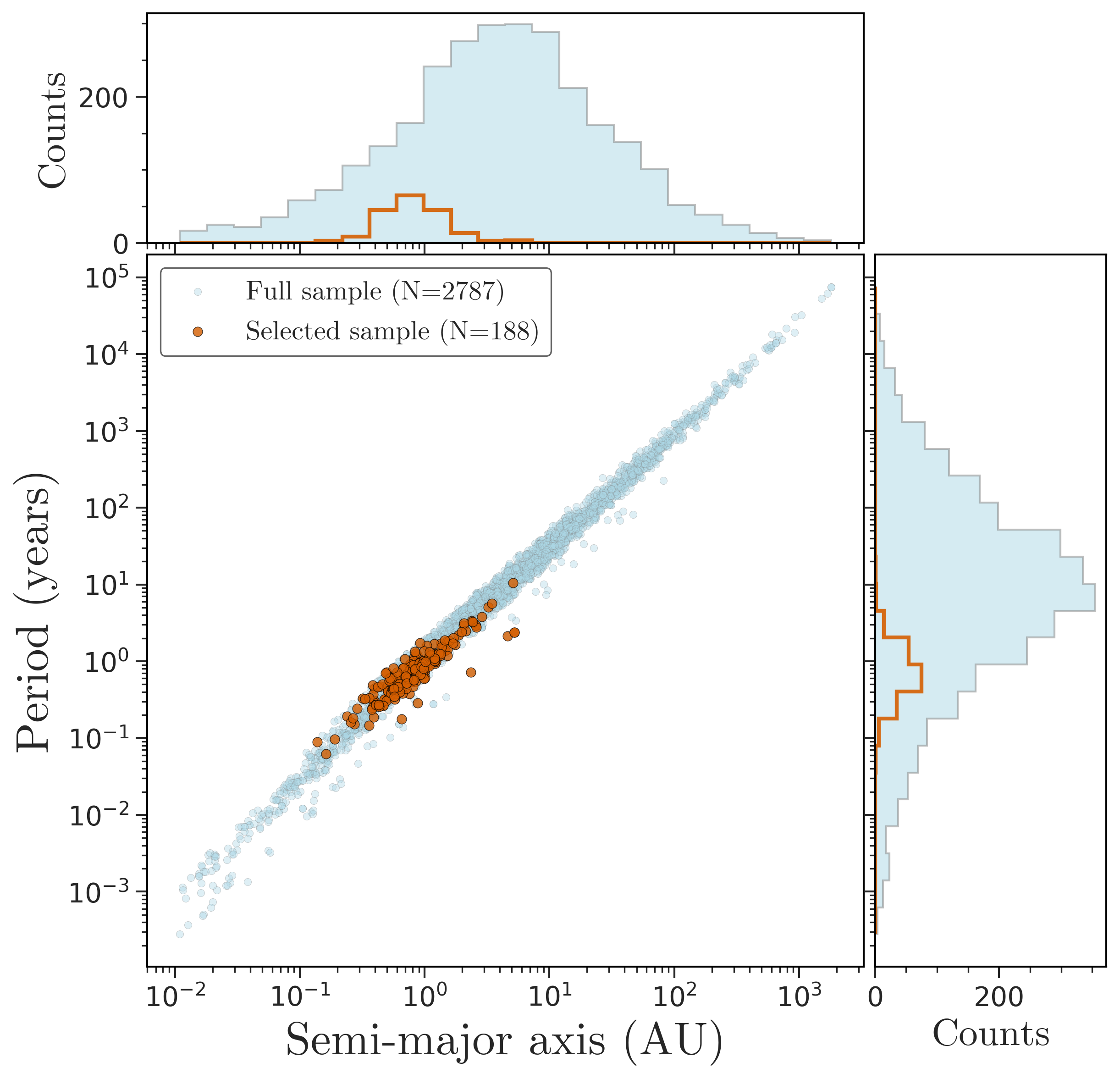}
         \caption{Period vs Semi-major axis for all binary lens events (light blue) with the sub-sample used for the cross section calculation highlighted in orange.}
         \label{fig:P_vs_a_full}
     \end{subfigure}
     
     \caption{Distribution of parameters for the full sample of 2787 binary lens microlensing events from \textit{PopSyCLE}}
     \label{fig:binary_param_dist}
\end{figure*}

\begin{figure*}[ht!]
    \begin{subfigure}[b]{0.48\linewidth}
        \centering
        \includegraphics[width=\linewidth]{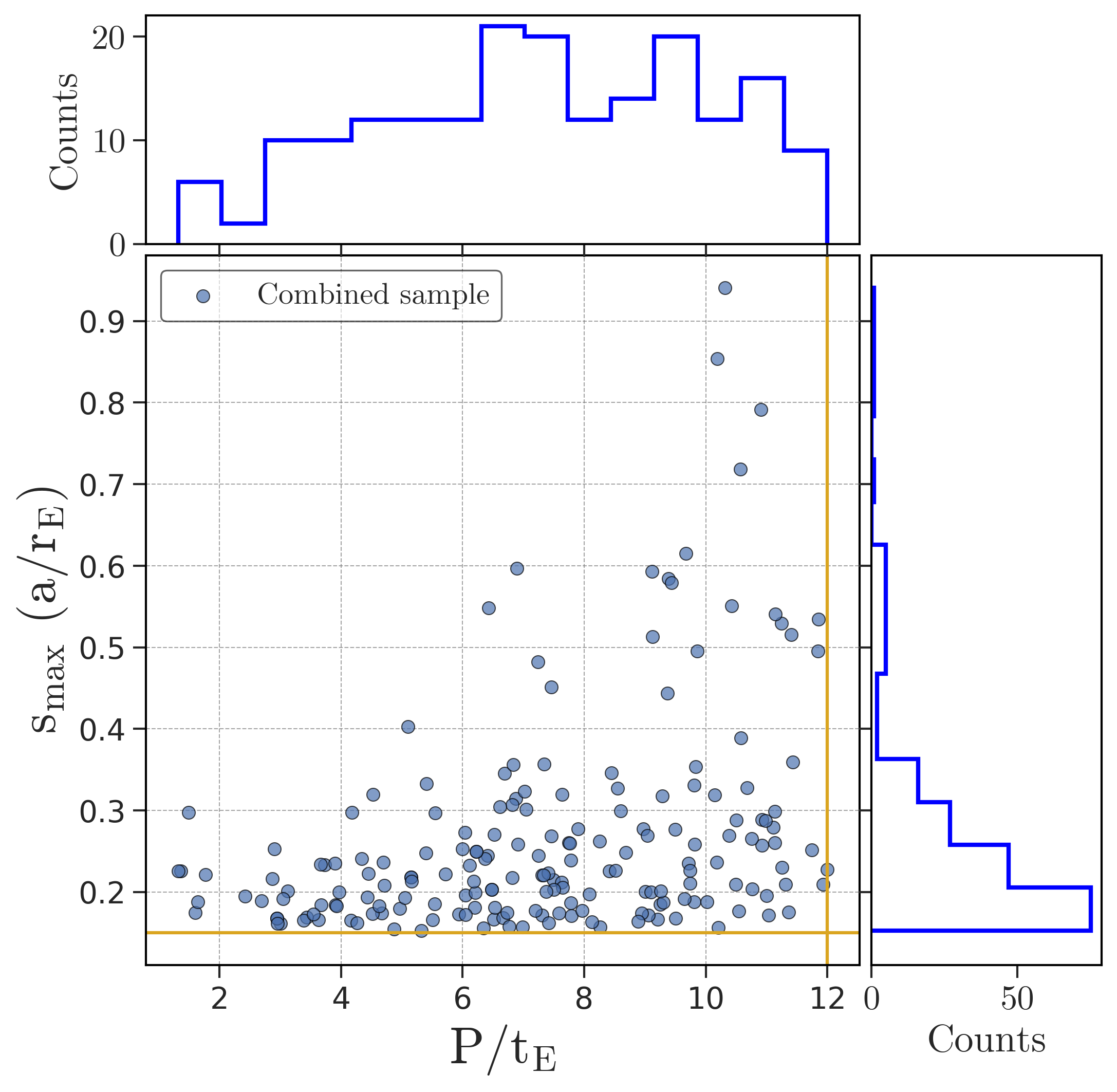}
        \captionsetup{margin={0cm, 0.3cm}}
        \caption{Distribution of $s_{\rm max}$ and $P/t_{\rm E}$ for the final sample of binary lens events.}
    \label{fig:real_sample}
    \end{subfigure}
    \hfill
    \begin{subfigure}[b]{0.43\linewidth}
        \includegraphics[width=\linewidth]{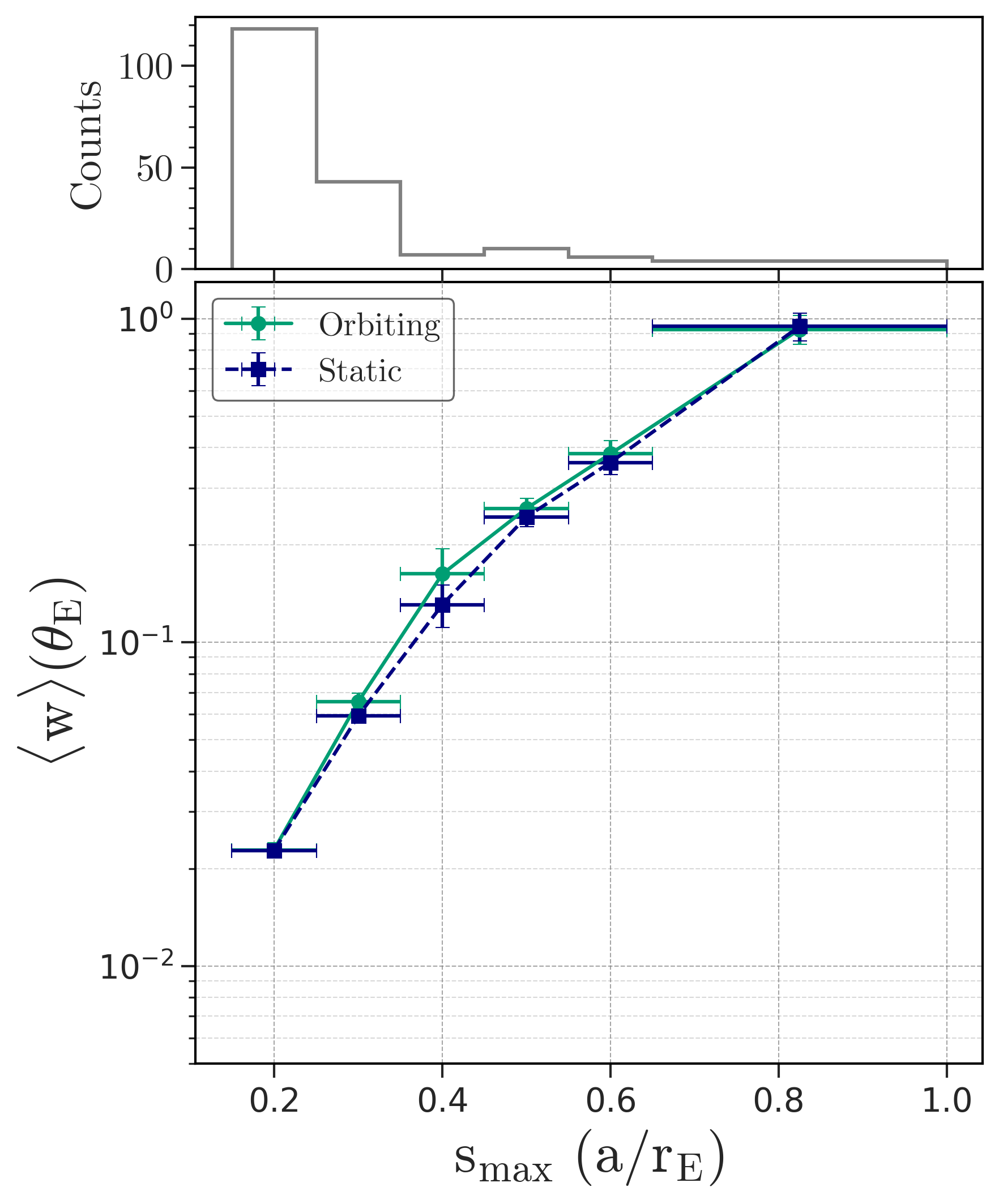}
        \captionsetup{margin={0.0cm, 0cm}}
        \caption{Caustic cross sections for the selected binary lens events assuming static (dark blue) and orbiting (green) lenses.}
        \label{fig:final_crs_results}
    \end{subfigure}
    \caption{Caustic cross section for simulated binary lens events produced by \textit{PopSyCLE} which have $P/t_{\rm E} < 12$ and $s_{\rm max} > 0.15$. Plot on the left is a scatter plot of $s_{\rm max}$ and $P/t_{\rm E}$ for all lenses in this sample. Orange lines show the cuts that have been applied. Right shows the cross section as a function of $s_{\rm max}$, where the systems have been binned in $s_{\rm max}$. The points are the mean cross section values of the binned systems, horizontal error bars represent bin widths, and vertical error bars show error in the mean value.}
    \label{fig:final_results}
\end{figure*} 

\begin{figure}
    \centering
    \includegraphics[width=0.98\linewidth]{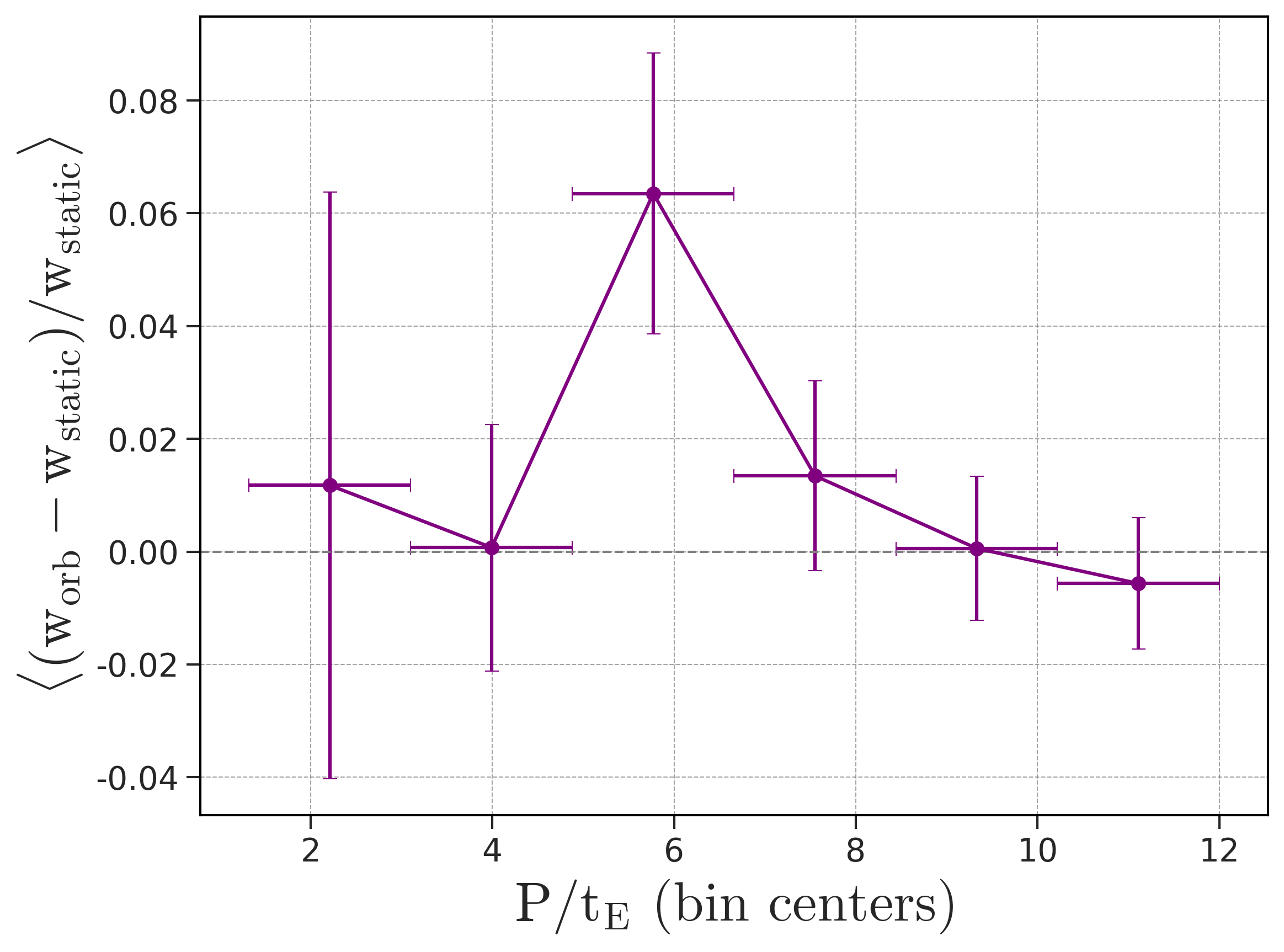}
    \caption{Mean fractional change in the cross section due to orbital motion effects as a function of $P/t_{\rm E}$. The systems have been binned in $P/t_{\rm E}$ and the x error bar represents the bin width. The y error bar is the error in the mean. }
    \label{fig:frac_dev_P_tE}
\end{figure}

Since the effect of orbital motion is most significant for small $P/t_{\rm E}$ values and larger caustic sizes, we further restricted the sample to events with $P/t_{\rm E} < 12$ and $s_{\rm max} > 0.15$. For events outside these bounds, we expect the deviation from the static assumption to be small (see Figs \ref{Fig:rotation_faceon_cross_sect}, \ref{fig:M_vs_s_plots}, \ref{fig:cross_sect_inc}). This restricted sample contains $\sim 60$ binary lens events. We then re-ran \textit{PopSyCLE} to obtain a larger and more statistically robust sample of lenses for the cross section calculation. We followed the same prescription as \citet{Abrams2025} for the simulations, but relaxed the observability criteria to $I_{\rm base} \leq 25$, and $|u_0| \leq 3$. This produced a larger ($\sim 1700$) sample of binary lens events which is statistically nearly identical to the first. We combined the two samples to produce a final sample that, after applying cuts to $P/t_{\rm E}$ and $s_{\rm max}$, contains 188 binary lens microlensing events, which corresponds to $\sim 7 \%$ of the full sample of binary lens events.

Fig \ref{fig:binary_param_dist} shows the distributions of various physical parameters of the full sample 2787 binary lens events from \textit{PopSyCLE} and the subset of events that passed our cuts for cross section calculation. Fig. \ref{fig:real_sample} shows the selected binary lens systems in the $s_{\rm max}$ vs $P/t_{\rm E}$ space. As expected from Fig. \ref{fig:M_vs_s_plots}, most binaries with small $P/t_{\rm E}$ values have small $s_{\rm max} \ (\lesssim 0.3)$. The number of systems with $s_{\rm max} > 0.4$, where we expect the effect of orbital motion to be most noticeable, is fairly small.

\subsection{Cross section calculation}
\label{sec5.2}

We calculated the caustic cross section for all 188 binary lens events selected using the procedure described above in two ways: by assuming the lenses are static during the event and by including orbital motion. We assumed circular orbits for simplicity of the calculations. For orbiting binaries, we averaged over 10 values of $\phi$ uniformly sampled between 0 and $\pi$ as before. In order to make a valid comparison between the static and orbiting cross sections, we must also account for the fact that, depending on the phase $\phi$, the projected separation $s$ during the event will typically be smaller than $s_{\rm max}$. Therefore, even when computing the cross sections for static binaries, we average over 10 $\phi$ values.  For each $\phi$ value, we determine $s$ at $t_0$ based on $s_{\rm max}$ and the inclination of the system.  This $s$ is assumed to be fixed during the event. This ensures that, for any binary system, the cross sections (both static and orbiting) we calculate is the average over all (projected) configurations in which we might find the binary at the time of the event.

Fig. \ref{fig:final_crs_results} shows the cross section as a function of $s_{\rm max}$, where the binary lens systems have been binned in $s_{\rm max}$ and the mean cross section in each bin has been plotted. The green curve shows the cross section for the orbiting case, and the dark blue curve shows the static approximation. There is very little difference between the two curves, with the orbiting case producing slightly larger cross sections for $s_{\rm max}$ between 0.3 and 0.6. The average cross section for this sample with orbital motion is (0.081 $\pm$ 0.011) $\theta_{\rm E}$, and the average static cross section is (0.077 $\pm$ 0.011) $\theta_{\rm E}$. The fractional difference between the average cross sections for the orbiting and static scenarios is $4.9\%$. Since the errors in mean values (and the deviation of values from the mean) of the static and orbiting cross section calculations are highly correlated, we cannot propagate those errors to the fractional change
\begin{figure*}[ht!]
    \begin{subfigure}[b]{0.48\linewidth}
        \centering
        \includegraphics[width=\linewidth]{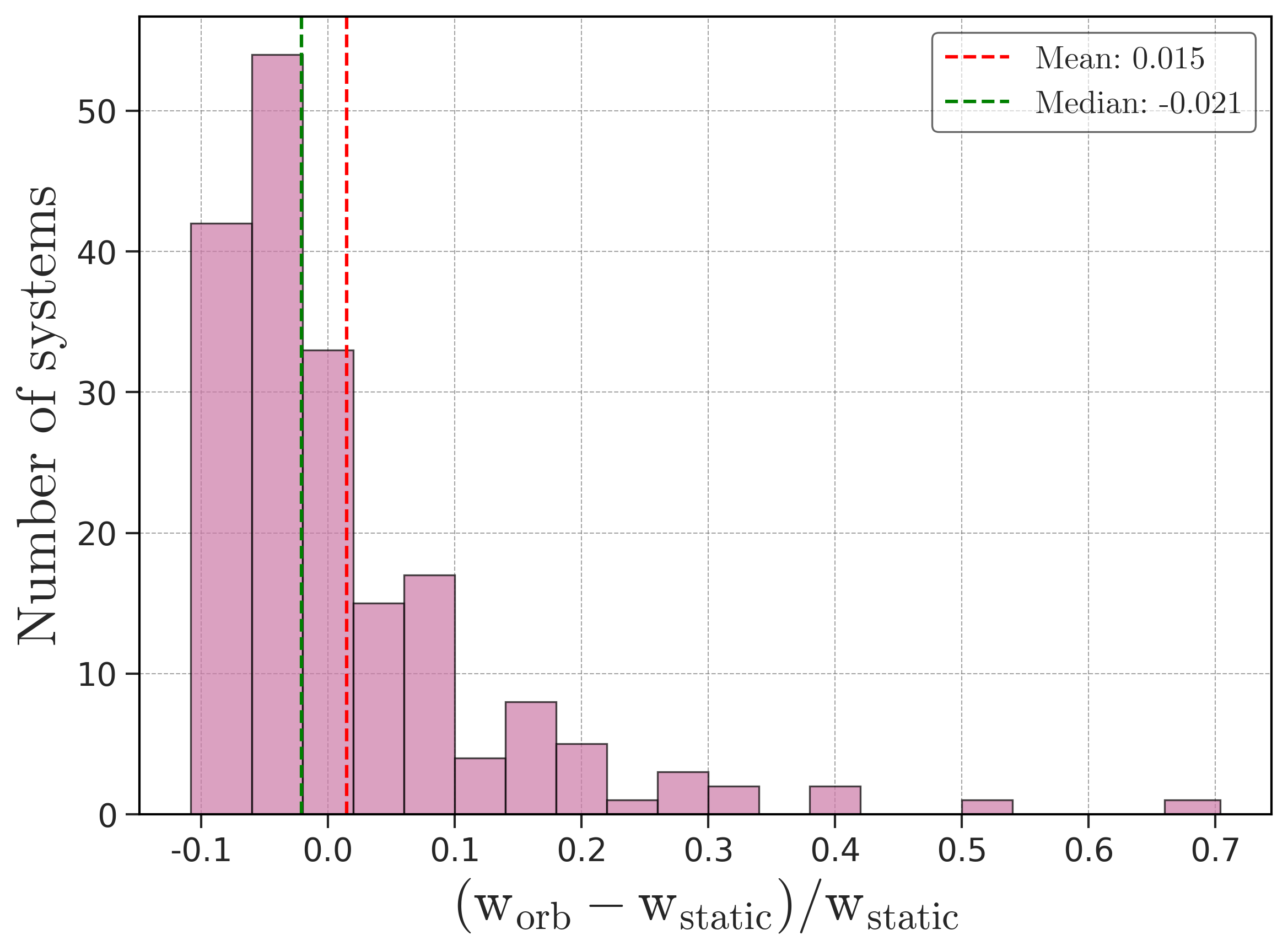}
        \captionsetup{margin={0.0cm, 0.4cm}}
        \caption{Distribution of fractional changes in cross section relative to a static binary. }
    \label{fig:frac_change_dist}
    \end{subfigure}
    \hfill
    \begin{subfigure}[b]{0.48\linewidth}
        \includegraphics[width=\linewidth]{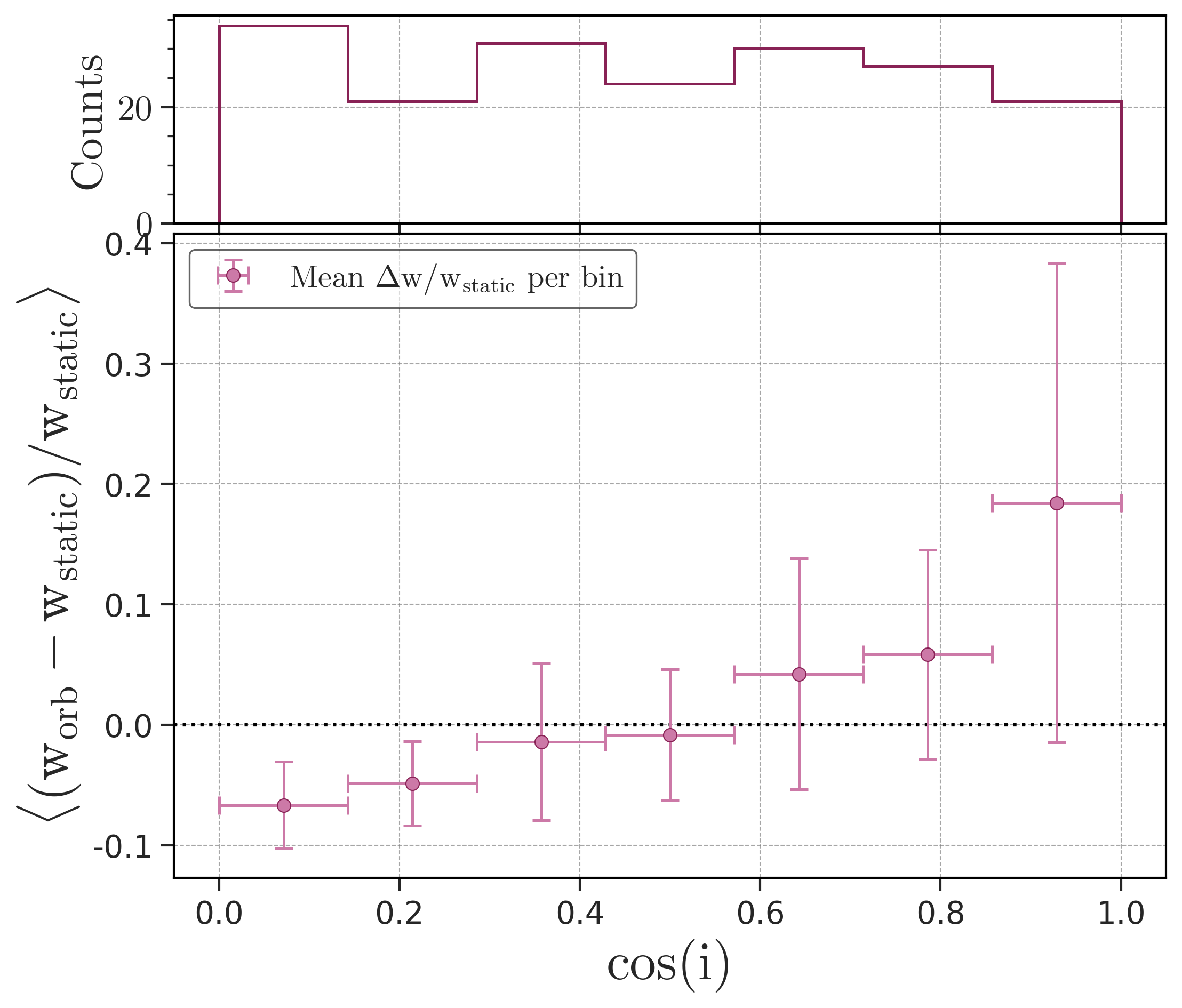}
        \captionsetup{margin={0.0cm, 0.4cm}}
        \caption{Fractional change as a function of inclination of the binary system.}
        \label{fig:frac_change_vs_inc}
    \end{subfigure}
    \caption{Fractional change in cross section due to orbital motion relative to a static binary lens for each system in Fig. \ref{fig:real_sample}. Plot on the left shows the differential distribution of the fractional change across the sample. Plot on the right shows the fractional change as a function of inclination, where the systems have been binned in cosine of the inclination angle. The horizontal error bars show the bin widths and the vertical error bars are the standard deviations of the systems in each bin.}
    \label{fig:frac_change_pop_plots}
\end{figure*}in mean cross section. Instead, we can also calculate the average fractional increase in cross section for a given system when orbital motion is included in the cross section calculation. This turns out to be $(1.5 \pm 0.8) \%$. Most of this deviation occurs at $s_{\rm max} \sim 0.4$ and $P/t_{\rm E} \sim 5 - 7$ (Fig. \ref{fig:frac_dev_P_tE}). This is because for a smaller $s_{\rm max}$, the caustic size is too small for there to be any significant deviation, and for a larger $s_{\rm max}$, the average $P/t_{\rm E}$ value is too large.

To understand why there is only a small ($\sim 2\%$) average fractional change in the cross section, we plotted the distribution of fractional change in the cross section of all systems (Fig. \ref{fig:frac_change_dist}). Most ($\sim 60 \%$) binary star systems have a small ($0 - 10\%$) \textit{decrease} in cross section relative to a static lens due to orbital motion. About $15 \%$ of the systems have a large ($> 20 \%$) \textit{increase} in cross section, and the rest have a smaller ($0 - 10 \%$) \textit{increase} in cross section relative to a static lens. Fig. \ref{fig:frac_change_vs_inc} shows the fractional change in the cross section as a function of the inclination of the binary system. Systems that are highly inclined (which are larger in number) have negative fractional changes, whereas systems that are closer to face on (smaller in number) have positive fractional changes. The magnitude of the change in cross section is relatively small for most systems, since most of them have small $s_{\rm max}$ or large $P/t_{\rm E}$ values. The increase in cross section relative to static binaries due to rapidly orbiting binary systems that are not highly inclined is nearly canceled out by the decrease in cross section due to the more edge-on binary systems, resulting in a net small average fractional increase in the cross section of orbiting binary stars relative to static ones in this sample.

For binary stars outside the selected sample, we expect the magnitude of change in the cross section to be much smaller (See Fig. \ref{Fig:rotation_faceon_cross_sect} and Fig. \ref{fig:cross_sect_inc}). If we assume that the average fractional change in cross section for all binaries outside the bounds of our sample is close to 0, then the mean fractional increase in cross section across all values of $s_{\rm max}$ and $P/t_{\rm E}$ is $\sim 0.1 \%$. Although we find that orbital motion does not shave a large impact on the average rate of caustic crossing binary events, we should still expect to see an over-representation of caustic crossing events due to more face-on binary star systems with orbital periods comparable to the Einstein ring crossing timescale relative to those caused by longer orbital period or edge-on binary star systems. If a sufficient number of caustic-crossing events are detected, one should find that such systems are more common than would be expected if binaries are static.

\subsection{Binaries with eccentric orbits}
\label{sec5.3}

All our computations of caustic cross sections assumed binary stars on circular orbits. The simple geometry of circular orbits allowed us to fully explore all the key effects of orbital motion while keeping the computations tractable in a reasonable amount of time. The investigation of this effect for binaries with eccentric orbits is beyond the scope of this paper. But we can make some qualitative arguments about what we expect to see for binaries with general Keplerian orbits based on the discussion of circular orbits above.

Consider a face-on elliptical orbit with a semi-major axis $a$ and eccentricity $e$. The time-averaged separation between the two stars is $r_{\rm mean} = a(1 + \frac{e^2}{2})$. If the orbital period of the binary system is comparable to the event timescale, then, the caustics will sweep out an area in the source plane as in the circular orbit case. However, for elliptical orbits, the separation between the two stars also changes with time. Let us compare the caustic cross section of this system when it is orbiting to when it is static and the binary stars have a separation equal to $r_{\rm mean}$. At the periapsis, the separation between the two stars is $ a(1 - e) < r_{\rm mean}$, resulting in a smaller caustic cross section relative to the static system (for close topology caustics). At the apoapsis, the separation is $a(1 + e) > r_{\rm mean}$, and hence the cross section should also be larger. Although at every value of the projected separation $s$, the rotation of the caustics in the source plane increases the cross section, at separations smaller than $r_{\rm mean}$ there will most likely be a net decrease in cross section relative to the static lens as observed in systems with inclined circular orbits. Since the stars spend a longer amount of time closer to the apoapsis than the periapsis (the median separation in time $r_{\rm median} > r_{\rm mean}$ for all $e$), on average, orbital motion should increase the caustic cross section of face-on binaries in elliptical orbits.

When the binary star system is inclined, the maximum projected separation between the stars in one orbit is generally smaller than that in a face-on orbit. This will result in a smaller enhancement in the cross section. Furthermore, it is not necessary for the stars to spend a longer amount of time close to this maximum projected separation, and for certain projections, they might spend longer at their smallest projected separation. Therefore, on average, the caustic cross sections of binaries on inclined elliptical orbits might be smaller than if they were considered to be static.

Ultimately, these arguments are qualitative and not rigorous. So, it is difficult to say anything more about the magnitude of the change in cross section without detailed calculations.

\subsection{Rate of caustic crossing binary events}
\label{sec5.4}

Finally, we can also estimate the average cross section for the entire sample of binary lenses from \textit{PopSyCLE}. We assume all of them to be static, since the correction due to orbital motion is small. Using the static lens cross sections calculated in Fig. \ref{fig:static_crs} and bilinearly interpolating in $s$ and $q$, we estimated cross sections for all 2800 binary lens events. We find the average cross section to be 0.423 $\pm$ 0.009. The ratio of caustic crossing events to binary lens events with $u_0 < 1$ is $0.423/2 = 0.21$. To find the rate of caustic crossing events relative to all microlensing events, we calculated the average cross section by weighting each cross section value by the binary fraction corresponding to the primary lens mass. We calculated binary fractions for different types of stars from Table 1 of \citet{Offner2023}. Since most triple systems exist as hierarchical triples with an inner close binary and a widely separated third companion (e.g. \citealt{Tokovinin:2014}), the close binary stars effectively act as a binary lens system. To account for this, we added half the triple and higher order fraction (THF) from the table in \citet{Offner2023} to the binary fraction we calculated.  We ignore higher-order ($N\ge 4$) systems. The mean binary lens fraction calculated this way for the \textit{PopSyCLE} sample is $30.4 \%$. The weighted average cross section is $0.125$. Therefore, the rate of caustic crossing events relative to all microlensing events with $u_0 < 1$ is $12.5/2 = (6.3 \pm 0.2)\%$. Binary event catalogs published by the OGLE and MACHO surveys have, for example, detected caustic crossing event rates of $4 \%$ in MACHO data \citep{Alcock2000a}, $2 \%$ in the OGLE III 2002-2003 season \citep{Jaroszynski2004}, $3.5 \%$ in the 2004 season \citep{Jaroszynski2006}, $1.6 \%$ in the 2005 season \citet{Skowron2007}, and $1.7 \%$ in the 2006-2008 season \citep{Jaroszynski2010}. The average caustic crossing rate for all OGLE III seasons (2002-2008) is $2.0 \%$. We calculated these rates by dividing the total number of caustic crossing binary events in these catalogs by the number of all microlensing events with $u_0 < 1$. Thus, the caustic crossing rates found by OGLE and MACHO are smaller than what we calculated, which could be due to some of the smaller caustic crossings being missed because of low cadence and irregular light curve coverage.  Alternatively, the properties of binaries could differ significantly from those of our simulated population.

\section{Conclusions}

Caustic crossings are the most distinctive signatures of binary lens microlensing events. They represent a significant portion of all known microlensing events. Roman is expected to have near perfect completeness of caustic crossing binary events. The rate of caustic crossing events is a metric that can be nearly independent of detection efficiency, and thus robustly used to infer the properties of the underlying population of stellar and compact object binary systems in the galaxy. Roman will be the first mission capable of probing the population of remnants and low-luminosity brown dwarfs and stars in binaries across large mass and galactic distance scales. 

In this work, we investigated the effect of lens orbital motion on the cross section of caustics. When a binary star system is more nearly face-on, in most cases, orbital motion causes the caustics to sweep out a larger area in the source plane. We found that when the period, $P$, of the binary system is on the order of the microlensing event timescale, $t_{\rm E}$, there is a large increase in the cross section for face-on binary lenses (See Fig. \ref{Fig:rotation_faceon_cross_sect}). However, if the binary star system is inclined to our line of sight, the average projected separation over one orbit between the stars is smaller than two times the semi-major axis of the binary. This results in a smaller enhancement of the cross section compared to face-on binaries, and for highly inclined orbits, can result in cross sections that are smaller than those for static binaries as well (See Fig. \ref{fig:cross_sect_inc} and Fig. \ref{fig:frac_change_vs_inc}).

We then simulated a realistic sample of $\sim 2800$ binary lens microlensing events (See Fig. \ref{fig:binary_param_dist}) and calculated the caustic cross section assuming static and orbiting lenses for a subset of this sample where the correction due to orbital motion is expected to be the largest (See Fig. \ref{fig:final_results}). The mean fractional increase in the cross section for a given system in the subset when orbital motion is considered in the cross section calculation is $(1.5 \pm 0.8) \%$. This small average change in cross section can be partly attributed to the cancellation between the positive fractional change due to more face-on systems and the negative fractional change due to highly inclined systems (See Fig. \ref{fig:frac_change_pop_plots}), and partly to the fact that there are not many binary star systems that have both a small enough $P/t_{\rm E}$ and large enough maximum projected separation $s_{\rm max}$ (See Fig. \ref{fig:M_vs_s_plots}). When considering the full sample of binary lenses, this value is $\sim 0.1 \%$. Therefore, the change in the overall rate of caustic crossing binary events due to orbital motion is likely to be completely negligible.

Although orbital motion does not significantly change the total rate of caustic crossing events, we should still expect to see face-on binary lenses with small $P/t_{\rm E}$ values over-represented in the sample of stellar binary lens events. These systems might be present in the known sample of caustic crossing binary events. However, the sample size of all known caustic crossing events is relatively small, and we would need some sort of constraint on the period of the binary system from either lens mass measurement or the detection of orbital motion effects in the light curve. \textit{Roman} will detect a much larger sample of caustic crossing binary events, and it would be a fruitful exercise to search for these systems in the \textit{Roman} data.

 The effect of orbital motion has been detected in the light curves of several caustic crossing events and has allowed constraints to be placed on the properties of the lens orbit (e.g., \citet{Albrow2000, An2002, Gaudi2008}). \citet{Penny2011b} calculated that for a continuous monitoring survey, $7 \%$ of all stellar binary lenses will show detectable orbital motion effects in their light curves. In the extreme case where $P \leq t_{\rm E}$, light curves may show repeating binary lens features due to the same magnification pattern features sweeping over the source multiple times. If such features are detected, they can be used to accurately determine the lens period (\citet{Penny2011a}). Even more exotic effects occur when the orbital speed of the binary lenses is a considerable fraction of the speed of light. \citet{Zheng2000} found that the behavior of the outer triangular caustics changes dramatically. The speed of these caustics can exceed the speed of light in this regime, and their size grows larger (and consequently, their magnification grows weaker) as the orbital speed increases.

Finally, we also calculated the average cross section for the entire \textit{PopSyCLE} sample of 2800 lenses and found the rate of caustic crossing events relative to all $u_0 < 1$ microlensing events to be $6.3 \pm 0.2\%$. This calculation can be used to constrain the multiplicity, mass ratio, and separation distribution of galactic binaries by matching the observed rate of caustic crossing events from the upcoming \textit{Roman} GBTDS. 
\label{conc}

\begin{acknowledgments}
We thank Natasha Abrams for providing simulated events from her \textit{PopSyCLE} run, as well as for her valuable guidance on running the code and interpreting its results. We are grateful to Amber Malpas for her assistance with the \textit{PopSyCLE} dataset, and to Radek Poleski for valuable inputs on using \textit{MulensModel}. We also thank the anonymous referee for helpful comments that improved the quality of this manuscript.

Funding for AM and BSG was provided by NASA through Nancy Grace Roman Space Telescope Project grant 80NSSC24M0022 and by the Thomas Jefferson Chair for Discovery and Space Exploration endowment.
\end{acknowledgments}

\begin{contribution}
AM performed all calculations and wrote the majority of the manuscript. BSG assisted with the interpretation of the results, suggested directions for further exploration, contributed to the text, and read and commented on the final manuscript. TAT conceived the idea for the paper, provided insight into the magnitude of the effects, read and commented on the manuscript, and suggested additional analyses.


\end{contribution}

%

\software{MulensModel, VBMicrolensing, Numpy, Matplotlib, Cursor AI tools (for help with writing code)}


\appendix

\section{Comparison with Baltz and Gondolo (2001)}
\label{secA1}
\begin{figure}[h!]
    \centering
    \includegraphics[width=0.5\linewidth]{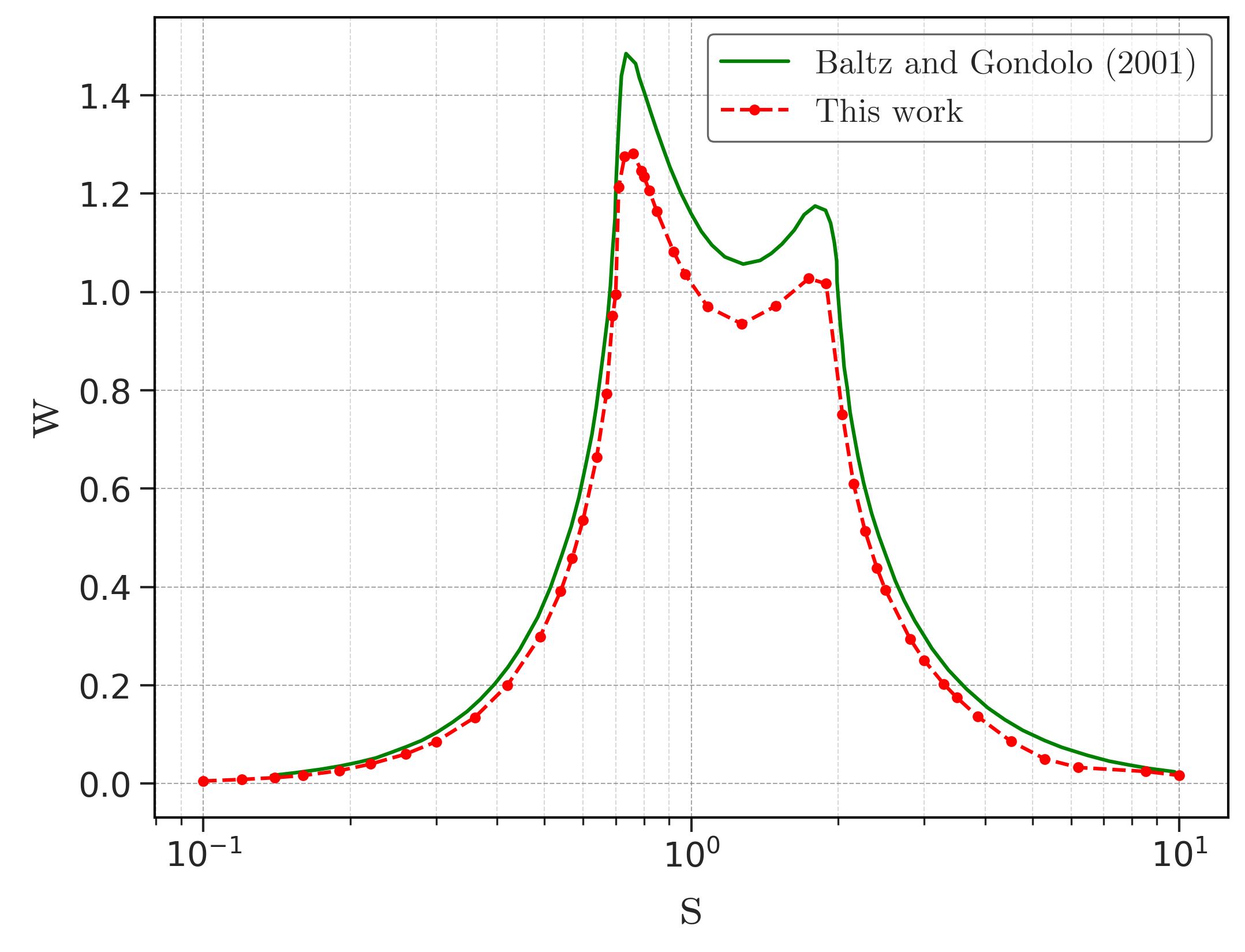}
    \caption{Comparison of static binary lens cross sections calculated by \citet{B&G2001} (green solid line) with the cross section calculated in this work (red dots and dashes). While the two curves are qualitatively similar, the cross sections we calculate are smaller than those in the Baltz and Gondolo paper for all $s$ values.}
    \label{fig:bng_comparison}
\end{figure}

\citet{B&G2001} define the ``width'' of a caustic, $w$ as (equation 19 in their paper) :
\begin{equation}
    w = \frac{l}{\pi} \ ,
\end{equation}
where $l$ is the length (perimeter) of the caustic curve. They calculate the length analytically by integrating the parametric form of caustics, and use this definition to then compute the rate of caustic crossing events $\Gamma_0$ as (equation 20 in the paper):
\begin{equation}
    \Gamma_0 = \Phi w 
\end{equation}
where $\Phi$ is the angle-averaged number flux of sources. Fig. \ref{fig:bng_comparison} shows the average ``width'' or cross section of caustics found by averaging $w$ over a flat $q$ distribution in the Baltz and Gondolo paper. While the qualitative shape of the curve they calculated agrees with what we found (including the location of the two peaks), their calculation produces larger values of cross sections for all $s$. This discrepancy can be explained as follows. The mean width of a closed \textit{convex} curve is equal to its perimeter divided by $\pi$ \citep{santalo2004integral}. However, this result cannot be extended to caustics which are \textit{concave} curves, as for any closed concave curve, we can always construct another curve with the same width but a larger perimeter (by making the edges more concave). In fact, the mean width of a concave curve is equal to the perimeter of the convex hull divided by $\pi$, where the convex hull is defined as the smallest convex set that encloses the curve. The perimeter of the {\it convex hull} of a concave curve is, by definition, smaller than the perimeter of the concave curve itself, and thus the mean width of a concave curve is smaller than its perimeter divided by $\pi$.  

The calculation of $\Gamma_0$ in equation 21 of the Baltz and Gondolo paper counts all entries to a caustic curve separately. Since concave curves can have multiple entries during a single event, this means that a caustic crossing event with two caustic entries will be counted twice in $\Gamma_0$. Therefore, $\Gamma_0$ is more appropriately the rate of caustic entries or half the rate of caustic peaks (in the light curve), and not the rate of caustic crossing microlensing events as we have defined in this paper. This difference in definition accounts for the larger cross sections found by Baltz and Gondolo.

\section{Explaining the smaller cross sections for orbiting face-on binaries in the resonant regime}
\label{A2}

\begin{figure*}[h!]
    \centering
    \begin{subfigure}[b]{0.45\linewidth}
        \includegraphics[width=\linewidth]{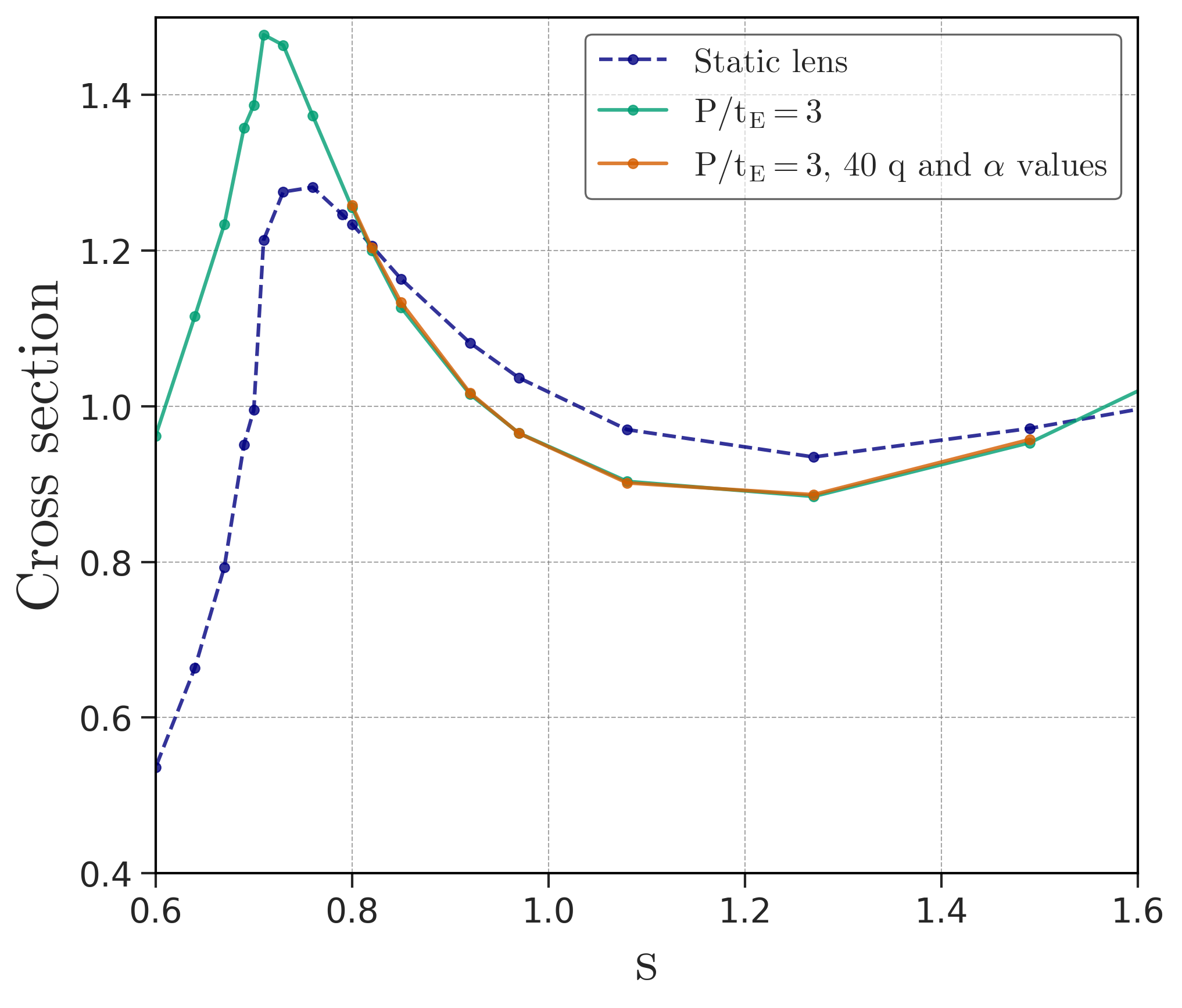}
        \caption{Cross section for a face-on orbiting lens with $P/t_{\rm E} = 3$ in the resonant regime }
        \label{fig:P3p0_crs}
    \end{subfigure}
    \hfill
    \begin{subfigure}[b]{0.45\linewidth}
        \includegraphics[width=\linewidth]{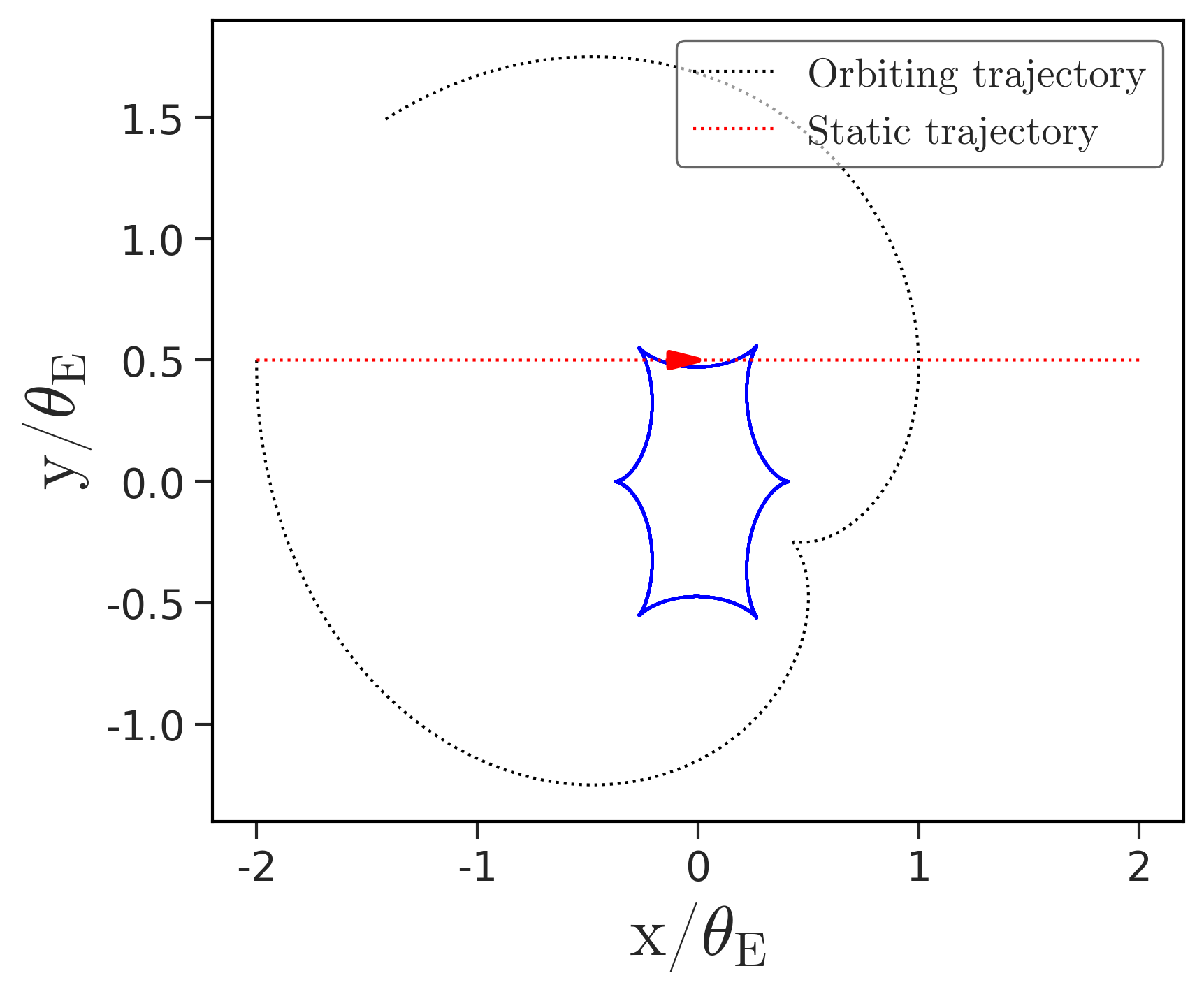}
        \caption{Static vs Orbiting trajectory for a face-on lens in the resonant regime}
        \label{fig:res_dip_explanation}
    \end{subfigure}
    \caption{(Left) Caustic cross sections for a static lens and an orbiting $P/t_{\rm E} = 3$ in the resonant caustic regime. The green curve is the original calculation, and the orange curve is the calculation with twice the number of $q$ and $\alpha$ values. (Right) Caustic structure (blue) and source trajectories for a binary lens in the resonant regime with $s = 1.1$ and $q = 0.9$. The red dotted line is the actual source trajectory, and the black dotted line shows the trajectory for an orbiting lens with $P/t_{\rm E} = 3$ in the rest frame of the lens stars.}
    \label{fig:P3p0_verify}
\end{figure*}

The cross section curves for $P/t_{\rm E} = 3$ and $P/t_{\rm E} = 6$ dip below the static cross section curve for $s > 0.8$ and then go back up when $s \gtrsim 1.5$ (Fig. \ref{fig:P3p0_crs}). This roughly corresponds to the resonant regime of caustics. To verify whether this is a real effect or if it is a numerical error due to imperfect sampling, we computed the cross section for this range of $s$ values for $P/t_{\rm E} = 3$ with twice the number of $q$ and $\alpha$ values. Fig. \ref{fig:P3p0_crs} shows the result of this computation. The cross section remains the same as our original calculation, confirming that this is a real effect. Fig \ref{fig:res_dip_explanation} shows the caustic of a binary lens in a resonant configuration. In red is a source trajectory that crosses the caustic, assuming that the lens is static. The black dotted line shows the same source trajectory if the lens was in a face-on orbit with $P/t_{\rm E} = 3$. The black trajectory is no longer caustic crossing since the caustic has rotated away by the time the source reaches it. In resonant caustics where the caustic is more elongated in one direction than the other, if the lens system does not complete one full orbit in the time it takes the source to cross the caustic, on average, the source sees a smaller caustic width. 

\section{The torus sampling strategy}
\label{A3}

\begin{figure*}[h!]
    \centering
    \includegraphics[width=0.5\linewidth]{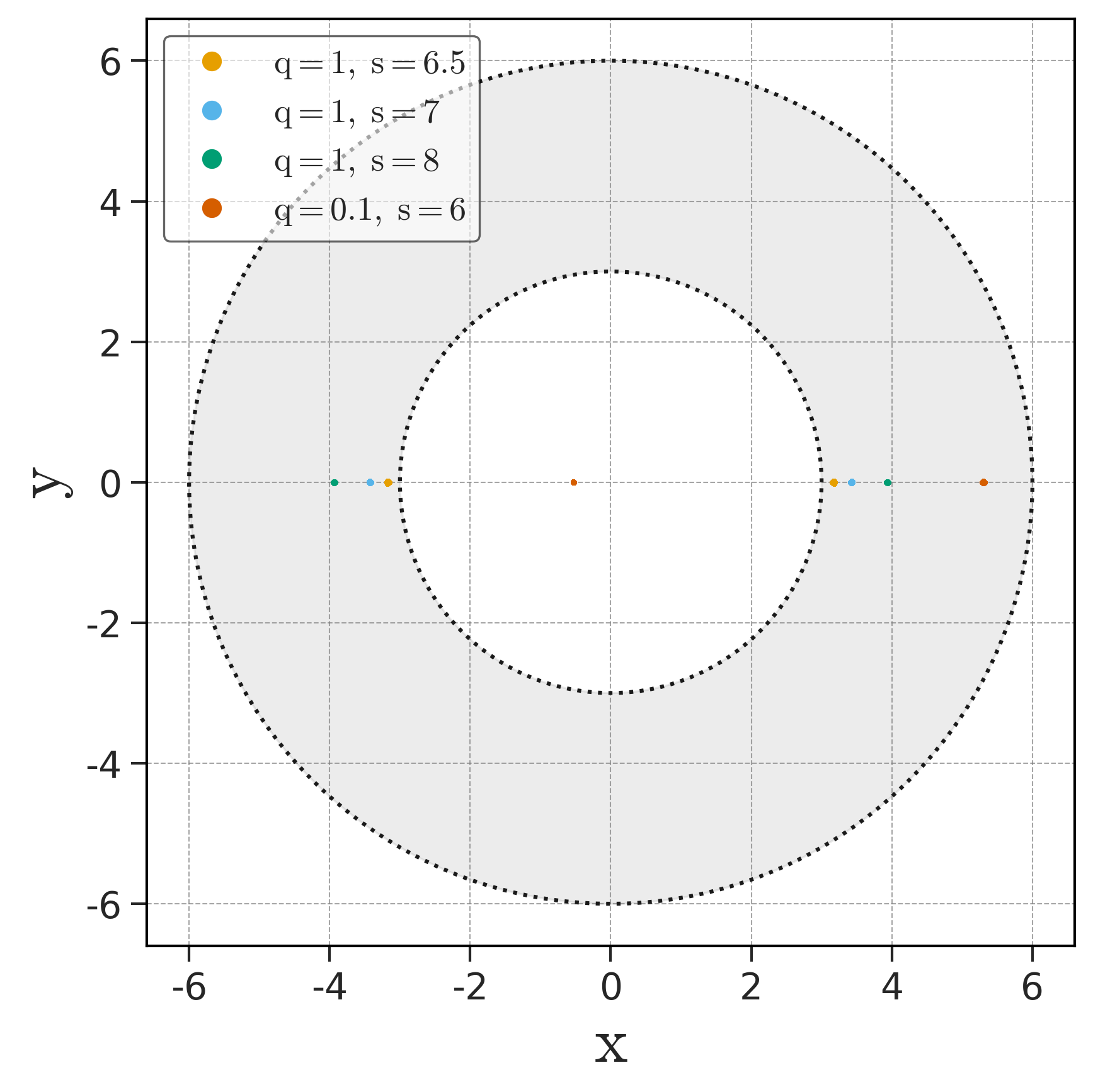}
    \caption{Caustics for $q = 1$, $s =$ 6.5, 7, and 8 and $q = 0.1, s = 6$. The shaded region represents the torus with $r_{in} = 3$ and $r_{out} = 6$. Caustics for equal mass binaries with $s > 6$ lie in the torus. For very unequal mass ratios and large $s$ (see $q = 0.1, \ s = 6$), one of the two diamond caustics is outside the torus region.}   
    \label{fig:torussampling}
\end{figure*}

 An equal mass binary in the wide topology produces two diamond caustics that are equidistant from the origin and are on either side of it. For large $s$, these caustics are widely separated. Fig \ref{fig:torussampling} shows caustics for $q =1$ and different values of $s > 6$. Both diamond caustics in this regime lie within a torus with $r_{in} = 3 \theta_{\rm E}$ and $r_{ext} = 6 \theta_{\rm E}$ (shaded region in Fig. \ref{fig:torussampling}). Sampling all points along a trajectory to calculate the width of these caustics will take a long time. Instead, we can only sample points on the trajectory that lie inside the torus. This is the approach we adopted for calculating cross sections of static and face on orbiting lenses with $s > 6$. For small mass ratios and large separations ($q = 0.1 \, s = 6$ in the figure), one of the two diamond caustics falls outside the torus and is missed by our calculation.

\clearpage

\bibliography{sample701}{}
\bibliographystyle{aasjournalv7}




\end{document}